\documentclass[aps,prl,reprint,superscriptaddress,amssymb,amsmath,floatfix,footinbib]{revtex4-1}
\pdfoutput=1
\usepackage{graphicx}
\usepackage{hyperref}

\newcommand{\YBCOx}{YBa\textsubscript{2}Cu\textsubscript{3}O\textsubscript{6+$x$}}
\newcommand{\YBCO}{YBa\textsubscript{2}Cu\textsubscript{3}O\textsubscript{6.67}}

\begin{document}

\title{Suppression of superconductivity by charge density wave order in \texorpdfstring{YBa\textsubscript{2}Cu\textsubscript{3}O\textsubscript{6.67}}{YBa2Cu3O6.67}} 

\author{Mark E. Barber}
\altaffiliation[Present address: ]{Department of Applied Physics and Geballe Laboratory for Advanced Materials, Stanford University, Stanford, CA 94305} 
\email{mebarber@stanford.edu}
\affiliation{Max Planck Institute for Chemical Physics of Solids, N{\"o}thnitzer Stra{\ss}e 40, 01187 Dresden, Germany}
\author{Hun-ho Kim}
\author{Toshinao Loew}
\affiliation{Max Planck Institute for Solid State Research, Heisenbergstra{\ss}e 1, 70569 Stuttgart, Germany}
\author{Matthieu Le Tacon}
\affiliation{Max Planck Institute for Solid State Research, Heisenbergstra{\ss}e 1, 70569 Stuttgart, Germany}
\affiliation{Karlsruhe Institute of Technology, Institute for Quantum Materials and Technologies, Hermann-von-Helmholtz-Platz 1, 76344 Eggenstein-Leopoldshafen, Germany}
\author{Matteo Minola}
\affiliation{Max Planck Institute for Solid State Research, Heisenbergstra{\ss}e 1, 70569 Stuttgart, Germany}
\author{Marcin Konczykowski}
\affiliation{Laboratoire des Solides Irradi{\'e}s, CEA/DRF/lRAMIS, Ecole Polytechnique, CNRS, Institut Polytechnique de Paris, F-91128 Palaiseau, France}
\author{Bernhard Keimer}
\affiliation{Max Planck Institute for Solid State Research, Heisenbergstra{\ss}e 1, 70569 Stuttgart, Germany}
\author{Andrew P. Mackenzie}
\affiliation{Max Planck Institute for Chemical Physics of Solids, N{\"o}thnitzer Stra{\ss}e 40, 01187 Dresden, Germany}
\affiliation{Scottish Universities Physics Alliance, School of Physics and Astronomy, University of St.\ Andrews, St.\ Andrews KY16 9SS, U.K.}
\author{Clifford W. Hicks} 
\email{hicks@cpfs.mpg.de}
\affiliation{Max Planck Institute for Chemical Physics of Solids, N{\"o}thnitzer Stra{\ss}e 40, 01187 Dresden, Germany}
\affiliation{School of Physics and Astronomy, University of Birmingham, Birmingham B15 2TT, U.K.}

\date{\today}

\begin{abstract}
Hole-doped cuprate superconductors show a ubiquitous tendency towards charge order.
Although onset of superconductivity is known to suppress charge order, there has not so far been a decisive demonstration of the reverse process, namely, the effect of charge order on superconductivity.
To gain such information, we report here the dependence of the critical temperature $T_{\mathrm{c}}$ of \YBCO{} on in-plane uniaxial stress up to 2~GPa.
At a compression of about 1~GPa along the $a$ axis, 3D-correlated charge density wave (3D CDW) order appears.
We find that $T_{\mathrm{c}}$ decreases steeply as the applied stress crosses 1~GPa, showing that the appearance of 3D CDW order strongly suppresses superconductivity.
Through the elastocaloric effect we resolve the heat capacity anomaly at $T_{\mathrm{c}}$, and find that it does not change drastically as the 3D CDW onsets, which shows that the condensation energy of the 3D CDW is considerably less than that of the superconductivity.
\end{abstract}

\maketitle

In the doping-temperature phase diagram of hole-doped cuprates there is always an antiferromagnetic phase at low doping and a dome of superconductivity at higher doping.
The proximity of superconductivity to antiferromagnetic order led early on to a hypothesis that antiferromagnetic fluctuations drive superconductivity, an idea that remains a bedrock of discussion about high-temperature superconductivity.
However, doping is only one possible axis for tuning.
Many hole-doped cuprates show short-range, quasi-two-dimensional CDW order when the doped hole density is around 1/8 per Cu site~\cite{Keimer15_Nature,Tabis17_PRB}.
In \YBCO{} (which has $p \approx 1/8$) a quasi-long-range, 3D-correlated charge order can be induced by magnetic field~\cite{Wu11_Nature,Gerber15_Science,Chang16_NatComm} or uniaxial stress~\cite{Kim18_Science}.
These forms of charge order are referred to as the 2D and 3D CDWs, respectively.
If hole-doped cuprate superconductivity is found to exist generally in proximity to both antiferromagnetism and charge order, along distinct tuning axes, the question then arises of whether fluctuations of both forms of order rather than of antiferromagnetism alone drive the superconductivity~\cite{CastroNeto01_PRB}.

To advance discussion, it is essential to understand how CDW order and superconductivity interact.
The onset of superconductivity suppresses the amplitude of the 2D CDW~\cite{Ghiringhelli12_Science,Croft14_PRB,Chang12_NatPhys}, and because competition is reciprocal the presence of the 2D CDW order must also weaken the superconductivity.
If, however, it has low spectral weight, for example if it condenses only in localised patches where the disorder configuration is favourable, then this effect might be quantitatively negligible, signalling that the susceptibility to CDW order might not be important in analysis of the superconductivity.
That the effect of the 2D CDW is not negligible is suggested by the facts that $T_{\mathrm{c}}$ of \YBCOx{} dips below trend for $p \sim 1/8$~\cite{Liang06_PRB}, and that under hydrostatic pressure $T_{\mathrm{c}}$ rises while 2D CDW order is suppressed~\cite{CyrChoiniere18_PRB, Souliou18_PRB}.
However, although this correlation is suggestive it remains in principle possible that the 2D CDW responds to, without substantially influencing, the superconductivity.
For example, there could be two quantum critical points under the superconducting dome~\cite{Grissonnanche14_NatComm}, yielding a natural dip in $T_{\mathrm{c}}$ between them.

Furthermore, until the relationship between the 2D and 3D CDWs is clarified, their interaction with the superconductivity should be regarded as separate lines of inquiry.
Unlike the 2D CDW, the strain-induced 3D CDW is suppressed completely when superconductivity onsets~\cite{Kim18_Science}.
(The field-induced 3D CDW and superconductivity are also mutually exclusive, at higher temperatures, but may coexist below $\approx$15~K~\cite{Kacmarcik18_PRL}.
A similar low-temperature coexistence region in the stress-temperature phase diagram is not excluded.)
However, this observation does not provide information on the effect of the 3D CDW on superconductivity.
In Ref.~\cite{Choi20_NatComm}, under an applied field the 3D CDW was found to reach maximum amplitude more quickly than the 2D CDW, suggesting that it suppresses superconductivity more effectively.
But the intrinsic inhomogeneity of the mixed state complicates analysis: because the 3D CDW is more rapidly suppressed by superconductivity, it may be more sharply confined to vortex cores.

\begin{figure}[tp]
 \centering
 \includegraphics{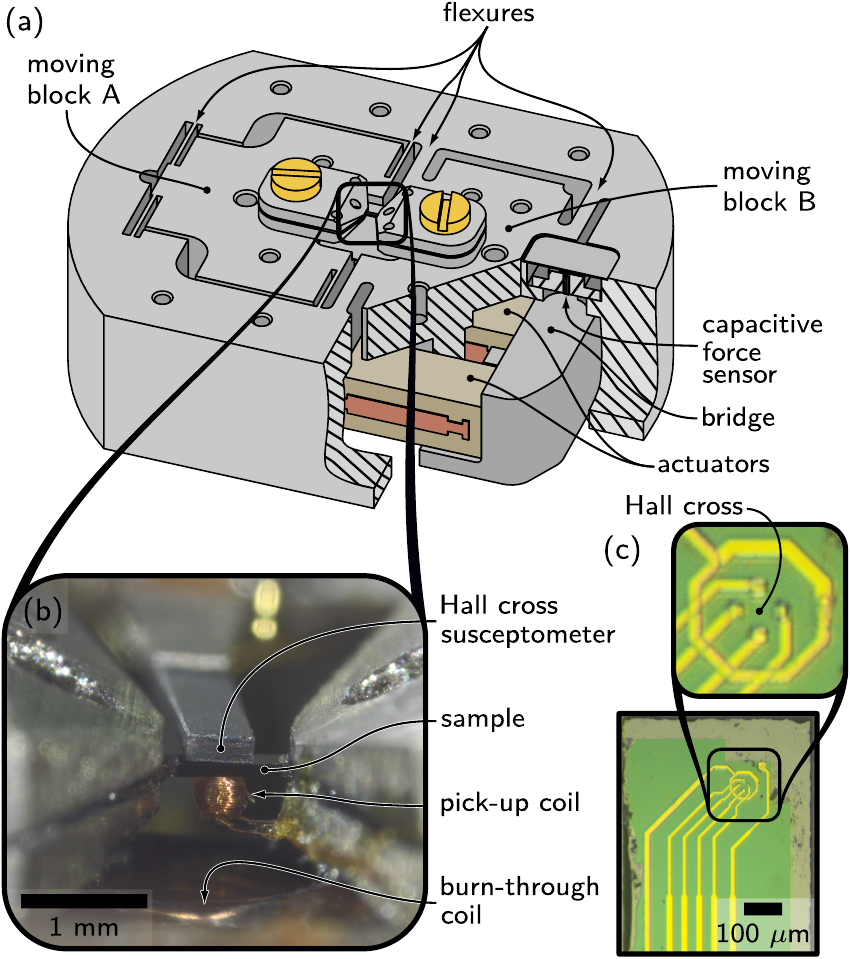}
 \caption{\label{fig1}(a)~Schematic of the piezoelectric-based uniaxial pressure cell employed here, including a cutaway showing the integrated force sensor and the placement of the actuators.
 The piezoelectric actuators drive motion of moving block A, through which force is applied to the sample.
 (b)~A micrograph of Sample 2, mounted for measurement.
 The Hall cross susceptometer rests on the upper surface of the sample, while pick-up and burn-through coils are placed beneath the sample.
 (c)~The tip of the microfabricated Hall probe susceptometer.
 The Hall cross itself is defined by proton irradiation, and is not visible.
}
\end{figure}

\begin{figure}[tp]
 \centering
 \includegraphics{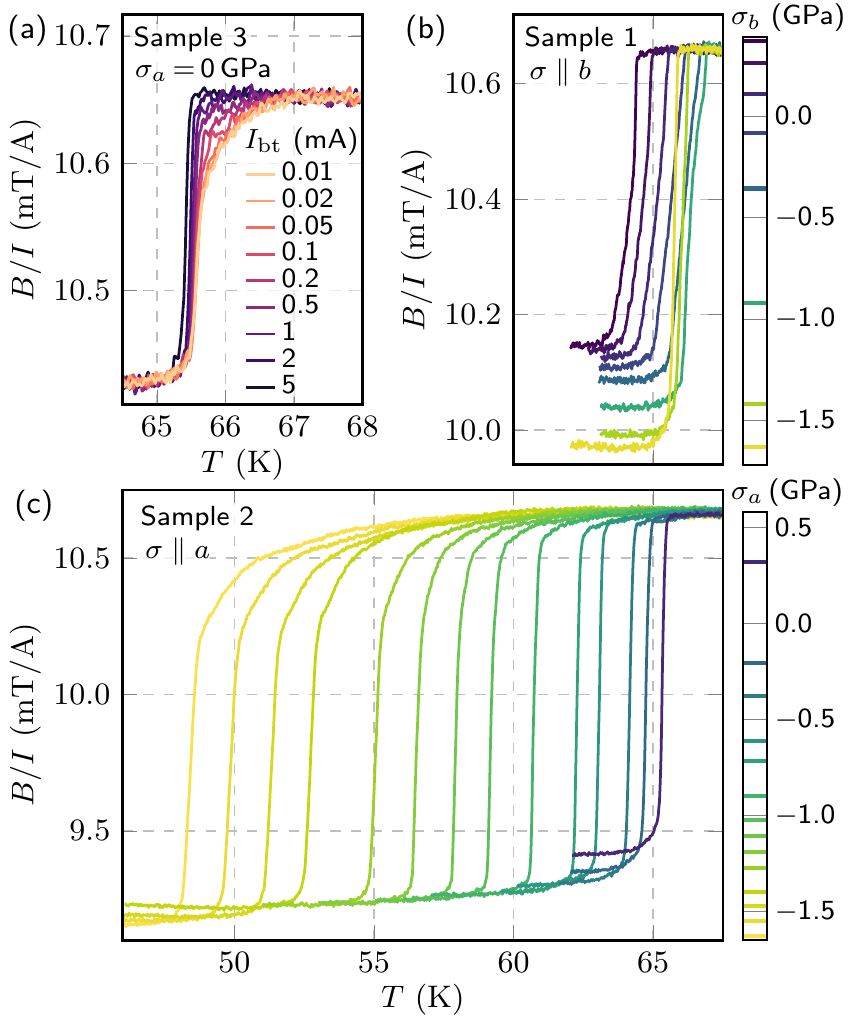}
 \caption{\label{fig2}(a)~The real part of the susceptometer response $B/I$ (field in the Hall cross divided by current in the excitation coil) for different currents in the burn-through coil $I_{\mathrm{bt}}$.
 The field from the excitation coil of the susceptometer at the sample surface was $\sim$25~$\mu$T at 211~Hz, while the burn-through field was applied at 20~kHz.
 5~mA in the burn-through coil corresponds to a field at the sample of $\sim$200~$\mu$T.
 (b)~$B/I$ versus $T$ for Sample 1 at various $b$-axis pressures $\sigma_{b}$.
 (c)~$B/I$ versus $T$ for Sample 2 at various $\sigma_{a}$.
}
\end{figure}

Here, we study the effect of charge order on the superconductivity of \YBCO{} through measurement of the uniaxial stress dependence of $T_{\mathrm{c}}$.
If the onset of 3D CDW order under $a$-axis stress $\sigma_{a}$ affects the superconductivity, an anomaly in $T_{\mathrm{c}}(\sigma_{a})$ is expected.
Considerable technical development was required to obtain enough precision for meaningful discussion.
We employ piezoelectric-based uniaxial stress apparatus [see Fig.~\ref{fig1}(a)], with which GPa-level uniaxial stresses are achievable by mounting beam-like samples with epoxy.
At these stresses, plastic deformation or fracture in the epoxy that partially relaxes the stress becomes a concern (see Fig.~S1), and therefore the apparatus incorporates a sensor of the applied force: the stress in the sample is the force divided by its cross-sectional area, regardless of the state of the epoxy.
In addition, to avoid averaging over long-length-scale stress and sample inhomogeneity, $T_{\mathrm{c}}$ was measured using a microfabricated susceptometer~\cite{Okazaki07_PRB}.
The sensor is a GaAs/AlGaAs 2DEG Hall cross with an active area of $10 \times 10$~$\mu$m$^2$ surrounded by a 70~$\mu$m-diameter excitation coil; see Fig.~\ref{fig1}(c).
The measured quantity is the field at the Hall cross divided by the excitation current, $B/I$.

Our setup also includes a ``burn-through'' coil, a large coil placed below the sample whose applied ac field aids flux motion.
As shown in Fig.~\ref{fig2}(a), the width of the superconducting transition, measured with the susceptometer, decreases from $\sim$1 to $\sim$0.1~K as the burn-through field is increased.
The applied fields are negligible in comparison with the upper critical field~\cite{Grissonnanche14_NatComm}, so this is an effect of inhomogeneous superconductivity.
The superconductivity of hole-doped cuprates has short-length-scale inhomogeneity~\cite{Gomes07_Nature}, so above the dominant $T_{\mathrm{c}}$ interacting patches of slightly higher $T_{\mathrm{c}}$ can form a Josephson network that hinders flux motion, giving the appearance of bulk superconductivity when the probing field is weak; this effect has been seen in Sr\textsubscript{2}RuO\textsubscript{4} with Ru inclusions~\cite{Kittaka10_PRB}.
Finally, there is a $\sim$300~$\mu$m-diameter pick-up coil beneath the sample, which in combination with the field from the burn-through coil can be used for a longer-length-scale measurement of $T_{\mathrm{c}}$.
This will be useful for comparison with elastocaloric effect measurements.

\begin{figure}[tp]
 \centering
 \includegraphics{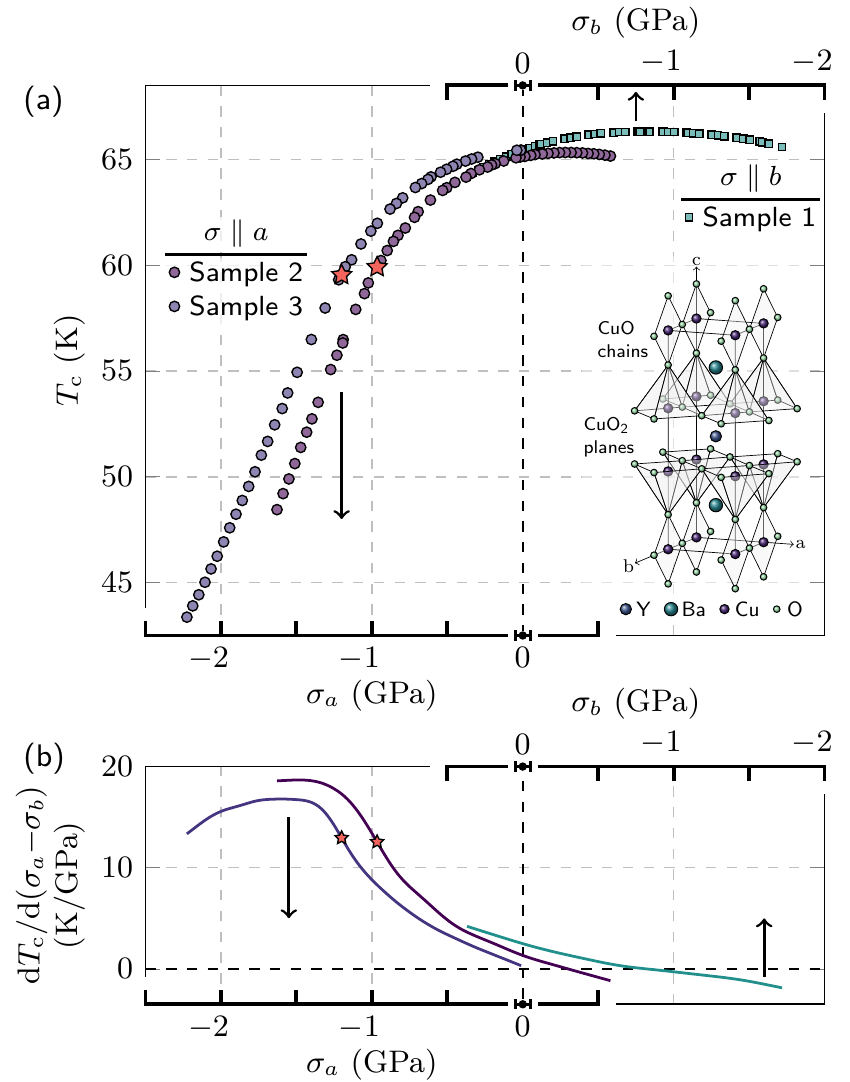}
 \caption{\label{fig3}(a)~Dependence of $T_{\mathrm{c}}$ on uniaxial stress, applied along the $b$ axis for Sample 1, and the $a$ axis for Samples 2 and 3.
 The stars mark the stresses, for Samples 2 and 3, where $|\mathrm{d}^2 T_{\mathrm{c}}/\mathrm{d}\sigma_{a}^2|$ is maximum, which we take as the onset stress $\sigma_{\mathrm{CDW}}$ for 3D CDW order.
 The error bars in the $\sigma$ axes indicate the error on the zero-force calibration.
 The inset shows the crystal structure of YBa\textsubscript{2}Cu\textsubscript{3}O\textsubscript{7}.
 (b)~Derivatives $\mathrm{d}T_{\mathrm{c}}/\mathrm{d}(\sigma_{a} - \sigma_{b})$, obtained from quintic smoothing splines of the data in the upper panel. 
}
\end{figure}

Uniaxial stress applied at room temperature can alter the oxygen ordering, and $T_\mathrm{c}$~\cite{Metzger93_PhysicaC,Fietz94_AIPConfProc}.
Here, large stress is applied only at temperatures well below 100~K.
We also periodically release the pressure and remeasure $T_\mathrm{c}$ to see if it was altered after the application of high pressures, but observe no such effects.
An example of the transition before and after the application of high pressure is shown in Fig.~S2.

Three samples were cut from the same single, detwinned crystal, Sample 1 for application of stress along the $b$ axis, and Samples 2 and 3 for stress along the $a$ axis.
Fig.~\ref{fig2}(b--c) shows susceptibility data from Samples 1 and 2 at various stresses.
The superconducting transitions remain narrow up to the highest stresses reached, even when $T_{\mathrm{c}}$ changes rapidly, showing that the probed regions have highly homogeneous $T_{\mathrm{c}}$.
Susceptibility data for Sample 3 are shown in Fig.~S3.

$T_{\mathrm{c}}$ data from all three samples, taking $T_{\mathrm{c}}$ as the point of maximum slope $\mathrm{d}(B/I)/\mathrm{d}T$, are shown together in Fig.~\ref{fig3}(a).
$\mathrm{d}T_{\mathrm{c}}/\mathrm{d}\sigma_{a}$ and $\mathrm{d}T_{\mathrm{c}}/\mathrm{d}\sigma_{b}$ have opposite signs, which shows that the in-plane orthorhombicity $b-a$ affects $T_{\mathrm{c}}$ more strongly than the unit cell area $ab$ or $c$-axis lattice constant (which also vary under uniaxial stress).
We therefore plot $T_{\mathrm{c}}(\sigma_{a})$ and $T_{\mathrm{c}}(\sigma_{b})$ together, with the $\sigma_{a}$ and $\sigma_{b}$ axes mirrored.

$b/a = 1.015$ in unstressed \YBCO{}~\cite{Krueger97_JSSC}, and for all three samples $T_{\mathrm{c}}$ initially increases when $b-a$ is decreased.
This is in agreement with thermal expansion data showing a reduction in $b-a$ below $T_{\mathrm{c}}$ in unstressed samples~\cite{Meingast91_PRL}.
We observe $\left. \mathrm{d}T_{\mathrm{c}}/\mathrm{d}\sigma_{b}\right|_{\sigma_{b} = 0} = -2.5$~K/GPa for Sample 1, and $\left. \mathrm{d}T_{\mathrm{c}}/\mathrm{d}\sigma_{a} \right|_{\sigma_{a} = 0} = +1.3$ and $+0.3$~K/GPa for Samples 2 and 3, respectively.
These are comparable to results on YBa\textsubscript{2}Cu\textsubscript{3}O\textsubscript{6.75} from Ref.~\cite{Welp94_JSupercond}: $\left. \mathrm{d}T_{\mathrm{c}}/\mathrm{d}\sigma_{b}\right|_{\sigma_{b} = 0} = -2.4$~K/GPa and $\left. \mathrm{d}T_{\mathrm{c}}/\mathrm{d}\sigma_{a}\right|_{\sigma_{a} = 0} = +1.6$~K/GPa.

However, the dominant feature in the data is a steep decrease in $T_{\mathrm{c}}$ for $\sigma_{a} < -1$~GPa.
This decrease is quasi-linear in applied stress; in Fig.~\ref{fig3}(b) we show the derivative $\mathrm{d}T_{\mathrm{c}}/\mathrm{d}\sigma_{a}$, which is seen to be approximately constant for $\sigma_{a} \lesssim -1.3$~GPa.

\begin{figure}[tp]
 \centering
 \includegraphics{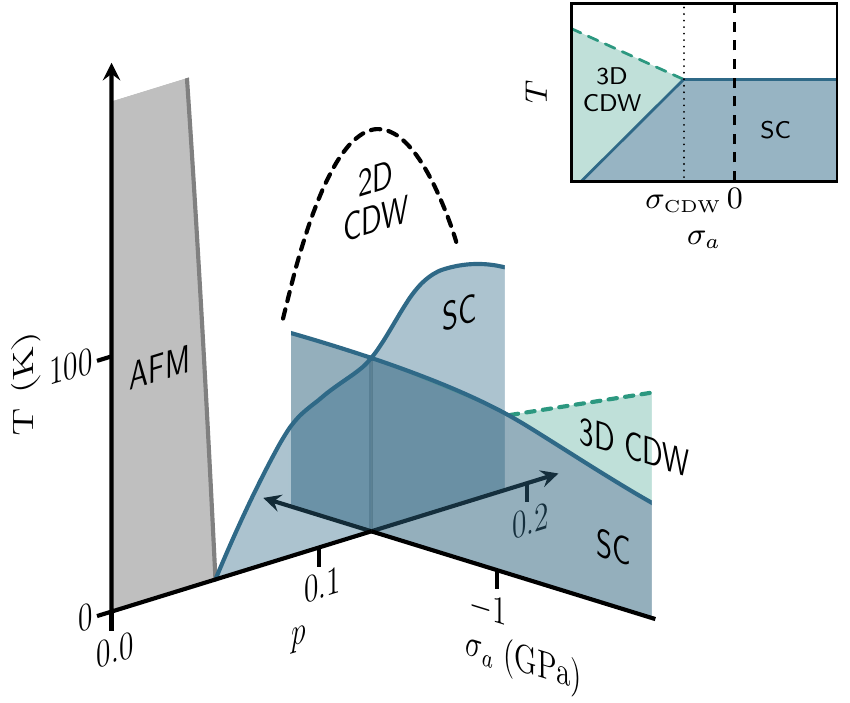}
 \caption{\label{fig4}Doping - uniaxial stress - temperature phase diagram of \YBCOx{}, as resolved in experiments so far.
 The antiferromagnetic, 2D CDW, and superconducting zero pressure phase boundaries are reproduced from \cite{Coneri10_PRB}, \cite{Blanco-Canosa14_PRB}, and \cite{Liang06_PRB}, respectively.
 The superconducting boundary in the stress-temperature plane is an average of Samples 2 and 3 from Fig.~\ref{fig3}(a).
 The onset temperature of the 3D CDW, for $|\sigma_{a}| > |\sigma_{\mathrm{CDW}}|$, has not yet been measured.
 Inset: schematic stress-temperature phase diagram for superconductivity suppressed by onset of uniaxial CDW order.
}
\end{figure}

To discuss this stress dependence of $T_{\mathrm{c}}$, we introduce a Ginzburg-Landau free energy functional for competing orders.
\begin{multline*}
F = \alpha_{d} \times (T - T_{\mathrm{c,0}})|\Delta_{d}|^2 + \beta_{d}|\Delta_{d}|^4 + g_{d}(\varepsilon_{\mathrm{o}} - \varepsilon_{\mathrm{o,0}})^2|\Delta_{d}|^2 + \\
\alpha_{\mathrm{C}} \times (T - T_{\mathrm{CDW,0}})|\Delta_{\mathrm{C}}|^2 + \beta_{\mathrm{C}}|\Delta_{\mathrm{C}}|^4 + g_{\mathrm{C}}(\varepsilon_{\mathrm{o}} - \varepsilon_{\mathrm{o},0})|\Delta_{\mathrm{C}}|^2 + \\
\lambda|\Delta_{\mathrm{C}}|^2|\Delta_{d}|^2 \, .
\end{multline*}
$\Delta_{d}$ and $\Delta_{\mathrm{C}}$ are the amplitudes of the superconducting and 3D CDW order parameters, respectively.
$\varepsilon_{\mathrm{o}} = \varepsilon_{xx} - \varepsilon_{yy}$ is the orthorhombic strain.
(We neglect terms for coupling to $c$-axis strain, $\varepsilon_{zz}$, and changes in unit cell area, $\varepsilon_{xx} + \varepsilon_{yy}$.)
The term $g_{d}(\varepsilon_{\mathrm{o}} - \varepsilon_{\mathrm{o,0}})^2|\Delta_{d}|^2$ represents direct coupling between the lattice and superconductivity, through the effect of lattice distortion on the pairing interaction.
While the lowest-order coupling between $d$-wave superconductivity and $\varepsilon_{\mathrm{o}}$ is quadratic in $\varepsilon_{\mathrm{o}}$, the 3D CDW is uniaxial~\cite{Chang16_NatComm}, and therefore couples linearly in $\varepsilon_{\mathrm{o}}$.
For both the superconductivity and CDW, the presence of the CuO chains introduces an offset $\varepsilon_{\mathrm{o},0} \neq 0$ (which in reality need not be the same for the CDW and superconductivity).
Conceptually, this is the strain at which the electronic interactions become tetragonally symmetric.

Because the 3D CDW is not seen in unstressed \YBCO{}, we must have $T_\mathrm{c,0} > T_\mathrm{CDW,0}$.
Because the superconductivity and 3D CDW compete, we must also have $\lambda > 0$.
In addition, the 3D CDW will choose the orientation that gives $g_{\mathrm{C}} (\varepsilon_{\mathrm{o}} - \varepsilon_{\mathrm{o},0}) < 0$.
In the inset of Fig.~\ref{fig4} we illustrate a schematic $\sigma$ - $T$ phase diagram where there is no direct coupling of the superconductivity to the lattice (that is, $g_{d} = 0$): $T_{\mathrm{c}}$ is unchanged until 3D CDW order onsets at $\sigma_{\mathrm{CDW}}$.
Competition then suppresses $T_{\mathrm{c}}$, and, due to the linear coupling between the 3D CDW and $\varepsilon_{\mathrm{o}}$, the suppression is stress-linear.

$T_{\mathrm{c}}(\sigma_a)$ has this qualitative form.
For $|\sigma_a| \lesssim 0.5$~GPa, $T_{\mathrm{c}}$ has a weak quadratic dependence on $\sigma_{a}$, that is potentially explainable through a small direct-interaction coefficient $g_d > 0$.
The crossover to a linear dependence for $\sigma_{a} \lesssim -1.3$~GPa can be straightforwardly explained by competition with a uniaxial order, the 3D CDW; obtaining this behaviour through direct-interaction terms alone would require an unlikely balance of higher-order terms.

Identifying $\sigma_{\mathrm{CDW}}$ in \YBCO{} as the maximum in $|\mathrm{d}^2 T_{\mathrm{c}}/\mathrm{d}\sigma_{a}^2|$, we obtain $\sigma_{\mathrm{CDW}} = -0.97$~GPa for Sample 2 and $-1.2$~GPa for Sample 3.
In the x-ray study of Ref.~\cite{Kim18_Science}, 3D CDW order was reported as onsetting at an $a$-axis strain between $-0.8$ and $-1.0 \cdot 10^{-2}$, corresponding to $\sigma_{\mathrm{CDW}}$ between $-1.3$ and $-1.6$~GPa~\cite{Lei93_PRB}.
Due to a nontrivial sample configuration there is some uncertainty in this strain scale; Bragg reflection data indicate that the 3D CDW could have onset at a strain as low as $-0.6$~$\cdot$~$10^{-2}$ (\textit{i.e.} $\sigma_{\mathrm{CDW}} = -1.0$~GPa).
Within uncertainty and possible sample-to-sample variation, $\sigma_\mathrm{CDW}$ from Ref.~\cite{Kim18_Science} agrees well with that found here.
The doping - uniaxial stress - temperature phase diagram of \YBCOx{}, as resolved so far, is shown in Fig.~\ref{fig4}.

In principle, strain-induced changes in the 2D CDW could also drive changes in $T_{\mathrm{c}}$.
However, recent x-ray data~\cite{Kim_unpublished} show that, in contrast to $T_{\mathrm{c}}$, the 2D CDW responds in a qualitatively symmetric way to uniaxial stress: $a$-axis compression strengthens the component with $\mathbf{q} \parallel \mathbf{b}$, and $b$-axis compression that with $\mathbf{q} \parallel \mathbf{a}$.
On the other hand, for stresses up to $\approx$1.7~GPa the 3D CDW appears only under $a$-axis compression.
Therefore the effects of the 3D CDW appear to dominate changes in $T_{\mathrm{c}}$.

To further investigate the relationship between 3D CDW order and superconductivity, we measure the heat capacity anomaly at $T_{\mathrm{c}}$ through the elastocaloric effect (ECE), the change in temperature in response to applied stress.
Competition means that the presence of the 3D CDW reduces the condensation energy of the superconductivity, and so for $\sigma_{a} < \sigma_{\mathrm{CDW}}$ a reduction in the heat capacity jump at $T_{\mathrm{c}}$, $\Delta C$, is expected.
The field-induced 3D CDW has nearly the same onset temperature as zero-field superconductivity~\cite{LeBoeuf13_NatPhys,Jang16_PNAS}, suggesting a close relationship between these phases.
If they were nearly degenerate, with similar condensation energies and entropy-temperature relations $S(T)$, the reduction in $\Delta C$ would be substantial.
This is possible: in one theoretical model, superconductivity and pair-density-wave order are found to be nearly degenerate~\cite{Corboz14_PRL}, and in experiments the stripe order of La\textsubscript{2$-x$}Ba\textsubscript{$x$}CuO\textsubscript{4} is seen to replace superconductivity with small changes in doping~\cite{Huecker11_PRB} or uniaxial stress~\cite{Guguchia20_PRL}, suggesting near-degeneracy.

\begin{figure}[tp]
 \centering
 \includegraphics{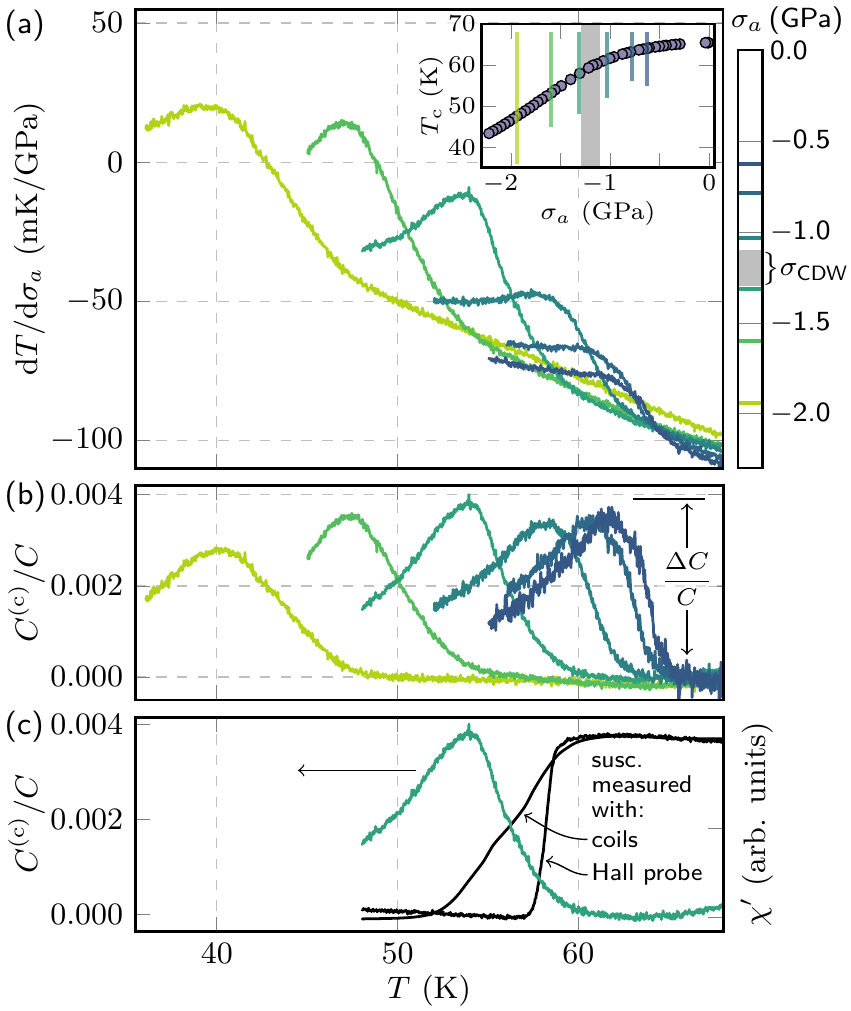}
 \caption{\label{fig5}(a)~Raw elastocaloric effect data from Sample 3.
 The $y$-axis is the change in temperature with stress under quasi-adiabatic conditions.
 The inset shows the locations of the temperature ramps on the $T_{\mathrm{c}}(\sigma_{a})$ curve.
 (b)~A normalisation of the data in panel (a), described in the text.
 $C^{\mathrm{(c)}}/C$ is the critical part of the heat capacity, associated with the superconducting transition, divided by the total heat capacity.
 (c)~Comparison of data at $\sigma_{a} = -$1.31~GPa with susceptibility measurements.
 The ECE measurements and susceptibility measurements with the coils both probe the entire exposed length of the sample, and so give nearly the same transition width.
}
\end{figure}

The elastocaloric effect arises due to stress-driven changes in entropy.
We employ the model of Ref.~\cite{Ikeda19_RSI}, in which the total heat capacity $C = C^{\mathrm{(nc)}} + C^{\mathrm{(c)}}$, where $C^{\mathrm{(c)}}$ is the heat capacity of critical electronic fluctuations associated with the onset of superconductivity and $C^{\mathrm{(nc)}}$ is the non-critical background.
$C^{\mathrm{(c)}}$ is taken to have the form $C^{\mathrm{(c)}}(T, \sigma) = C^{\mathrm{(c)}}(T - T_{\mathrm{c}}(\sigma))$.
In this model, and in the adiabatic limit,
\begin{equation*}
\frac{\mathrm{d} T}{\mathrm{d}\sigma} = \frac{C^{\mathrm{(c)}}}{C} \frac{\mathrm{d} T_{\mathrm{c}}}{\mathrm{d}\sigma} + \frac{\mathrm{d}T^{\mathrm{(nc)}}}{\mathrm{d}\sigma},
\end{equation*}
where $\mathrm{d} T^{\mathrm{(nc)}}/\mathrm{d}\sigma$ is the contribution from non-critical changes in entropy~\cite{Ikeda19_RSI}.
Measurements were performed on Sample 3, through the addition of a thermocouple to the sample setup.
The measurement frequency was 23~Hz, at which the thermal diffusion length is $\sim$500~$\mu$m in our temperature range (Ref.~\cite{Zhang17_PNAS}, and Fig.~S4), meaning that this measurement averages over essentially the entire exposed portion of the sample.

Raw data at various $\sigma_{a}$ are shown in Fig.~\ref{fig5}(a).
In Fig.~\ref{fig5}(b), an approximate normalisation is obtained by subtracting a common slope from all the curves, as an estimate for $\mathrm{d} T^{\mathrm{(nc)}}/\mathrm{d}\sigma$, then dividing each curve by $\mathrm{d} T_{\mathrm{c}}/\mathrm{d}\sigma_{a}$.
(Curves are also divided by a correction factor of $\approx$0.8, to account for the fact that measurements were not fully in the adiabatic limit; see Supplemental Materials for details.)
In panel (c), data at $\sigma_{a} = -1.31$~GPa are compared with the susceptibility as measured with the combination of pick-up and burn-through coils.
The transition widths match, confirming that the feature seen in the ECE is the superconducting transition.
$\Delta C$ at $T_{\mathrm{c}}$ is found to be $\sim$0.4\% of the total heat capacity [See Fig.~\ref{fig5}(b)], which is in good agreement with that obtained through direct measurement on unstressed samples, 0.3--0.5\%~\cite{Loram93_PRL}.
This demonstrates the capability of ECE measurements to resolve tiny heat capacity anomalies, even with inhomogeneity broadening.

We find that any change in $\Delta C$ across $\sigma_{\mathrm{CDW}}$ is small, meaning that although the presence of 3D CDW order strongly suppresses $T_{\mathrm{c}}$, its condensation energy is not close to that of superconductivity.
We do not resolve elastocaloric anomalies that could indicate onset of 3D CDW order, possibly because our data do not extend to high enough temperature, or because the stress dependence of the transition temperature is too small.

Identifying the phase boundary where the 3D CDW onsets is an important future task to better understand the relevance of charge order for superconductivity.
By showing that the onset of the 3D CDW substantially weakens the superconductivity, our result here shows that it is not a low-spectral-weight curiosity, but a phase that is closely intertwined with the superconductivity.\\

\begin{acknowledgments}
We thank Erez Berg, Ted Forgan, Stephen Hayden, J\"{o}rg Schmalian, John Tranquada, and Steve Kivelson for useful discussions.
M.K. thanks Vincent Mosser for his contributions to development of the Hall cross susceptometers.
We acknowledge the financial support of the Max Planck Society.
M.K. acknowledges financial support from ANR LABEX grant number ANR-10-LABX-0039-PALM.
\end{acknowledgments}


\begin{thebibliography}{36}%
\makeatletter
\providecommand \@ifxundefined [1]{%
 \@ifx{#1\undefined}
}%
\providecommand \@ifnum [1]{%
 \ifnum #1\expandafter \@firstoftwo
 \else \expandafter \@secondoftwo
 \fi
}%
\providecommand \@ifx [1]{%
 \ifx #1\expandafter \@firstoftwo
 \else \expandafter \@secondoftwo
 \fi
}%
\providecommand \natexlab [1]{#1}%
\providecommand \enquote  [1]{``#1''}%
\providecommand \bibnamefont  [1]{#1}%
\providecommand \bibfnamefont [1]{#1}%
\providecommand \citenamefont [1]{#1}%
\providecommand \href@noop [0]{\@secondoftwo}%
\providecommand \href [0]{\begingroup \@sanitize@url \@href}%
\providecommand \@href[1]{\@@startlink{#1}\@@href}%
\providecommand \@@href[1]{\endgroup#1\@@endlink}%
\providecommand \@sanitize@url [0]{\catcode `\\12\catcode `\$12\catcode
  `\&12\catcode `\#12\catcode `\^12\catcode `\_12\catcode `\%12\relax}%
\providecommand \@@startlink[1]{}%
\providecommand \@@endlink[0]{}%
\providecommand \url  [0]{\begingroup\@sanitize@url \@url }%
\providecommand \@url [1]{\endgroup\@href {#1}{\urlprefix }}%
\providecommand \urlprefix  [0]{URL }%
\providecommand \Eprint [0]{\href }%
\providecommand \doibase [0]{http://dx.doi.org/}%
\providecommand \selectlanguage [0]{\@gobble}%
\providecommand \bibinfo  [0]{\@secondoftwo}%
\providecommand \bibfield  [0]{\@secondoftwo}%
\providecommand \translation [1]{[#1]}%
\providecommand \BibitemOpen [0]{}%
\providecommand \bibitemStop [0]{}%
\providecommand \bibitemNoStop [0]{.\EOS\space}%
\providecommand \EOS [0]{\spacefactor3000\relax}%
\providecommand \BibitemShut  [1]{\csname bibitem#1\endcsname}%
\let\auto@bib@innerbib\@empty
%</preamble>
\bibitem [{\citenamefont {Keimer}\ \emph {et~al.}(2015)\citenamefont {Keimer},
  \citenamefont {Kivelson}, \citenamefont {Norman}, \citenamefont {Uchida},\
  and\ \citenamefont {Zaanen}}]{Keimer15_Nature}%
  \BibitemOpen
  \bibfield  {author} {\bibinfo {author} {\bibfnamefont {B.}~\bibnamefont
  {Keimer}}, \bibinfo {author} {\bibfnamefont {S.~A.}\ \bibnamefont
  {Kivelson}}, \bibinfo {author} {\bibfnamefont {M.~R.}\ \bibnamefont
  {Norman}}, \bibinfo {author} {\bibfnamefont {S.}~\bibnamefont {Uchida}}, \
  and\ \bibinfo {author} {\bibfnamefont {J.}~\bibnamefont {Zaanen}},\
  }\bibfield  {title} {\enquote {\bibinfo {title} {{From quantum matter to
  high-temperature superconductivity in copper oxides}},}\ }\href {\doibase
  10.1038/nature14165} {\bibfield  {journal} {\bibinfo  {journal} {Nature}\
  }\textbf {\bibinfo {volume} {518}},\ \bibinfo {pages} {179--186} (\bibinfo
  {year} {2015})}\BibitemShut {NoStop}%
\bibitem [{\citenamefont {Tabis}\ \emph {et~al.}(2017)\citenamefont {Tabis},
  \citenamefont {Yu}, \citenamefont {Bialo}, \citenamefont {Bluschke},
  \citenamefont {Kolodziej}, \citenamefont {Kozlowski}, \citenamefont
  {Blackburn}, \citenamefont {Sen}, \citenamefont {Forgan}, \citenamefont
  {Zimmermann}, \citenamefont {Tang}, \citenamefont {Weschke}, \citenamefont
  {Vignolle}, \citenamefont {Hepting}, \citenamefont {Gretarsson},
  \citenamefont {Sutarto}, \citenamefont {He}, \citenamefont {Le~Tacon},
  \citenamefont {Bari{\v{s}}i{\'{c}}}, \citenamefont {Yu},\ and\ \citenamefont
  {Greven}}]{Tabis17_PRB}%
  \BibitemOpen
  \bibfield  {author} {\bibinfo {author} {\bibfnamefont {W.}~\bibnamefont
  {Tabis}}, \bibinfo {author} {\bibfnamefont {B.}~\bibnamefont {Yu}}, \bibinfo
  {author} {\bibfnamefont {I.}~\bibnamefont {Bialo}}, \bibinfo {author}
  {\bibfnamefont {M.}~\bibnamefont {Bluschke}}, \bibinfo {author}
  {\bibfnamefont {T.}~\bibnamefont {Kolodziej}}, \bibinfo {author}
  {\bibfnamefont {A.}~\bibnamefont {Kozlowski}}, \bibinfo {author}
  {\bibfnamefont {E.}~\bibnamefont {Blackburn}}, \bibinfo {author}
  {\bibfnamefont {K.}~\bibnamefont {Sen}}, \bibinfo {author} {\bibfnamefont
  {E.~M.}\ \bibnamefont {Forgan}}, \bibinfo {author} {\bibfnamefont {M.~v.}\
  \bibnamefont {Zimmermann}}, \bibinfo {author} {\bibfnamefont
  {Y.}~\bibnamefont {Tang}}, \bibinfo {author} {\bibfnamefont {E.}~\bibnamefont
  {Weschke}}, \bibinfo {author} {\bibfnamefont {B.}~\bibnamefont {Vignolle}},
  \bibinfo {author} {\bibfnamefont {M.}~\bibnamefont {Hepting}}, \bibinfo
  {author} {\bibfnamefont {H.}~\bibnamefont {Gretarsson}}, \bibinfo {author}
  {\bibfnamefont {R.}~\bibnamefont {Sutarto}}, \bibinfo {author} {\bibfnamefont
  {F.}~\bibnamefont {He}}, \bibinfo {author} {\bibfnamefont {M.}~\bibnamefont
  {Le~Tacon}}, \bibinfo {author} {\bibfnamefont {N.}~\bibnamefont
  {Bari{\v{s}}i{\'{c}}}}, \bibinfo {author} {\bibfnamefont {G.}~\bibnamefont
  {Yu}}, \ and\ \bibinfo {author} {\bibfnamefont {M.}~\bibnamefont {Greven}},\
  }\bibfield  {title} {\enquote {\bibinfo {title} {{Synchrotron x-ray
  scattering study of charge-density-wave order in
  ${\mathrm{HgBa}}_{2}{\mathrm{CuO}}_{4+\ensuremath{\delta}}$}},}\ }\href
  {\doibase 10.1103/PhysRevB.96.134510} {\bibfield  {journal} {\bibinfo
  {journal} {Phys. Rev. B}\ }\textbf {\bibinfo {volume} {96}},\ \bibinfo
  {pages} {134510} (\bibinfo {year} {2017})}\BibitemShut {NoStop}%
\bibitem [{\citenamefont {Wu}\ \emph {et~al.}(2011)\citenamefont {Wu},
  \citenamefont {Mayaffre}, \citenamefont {Kr{\"{a}}mer}, \citenamefont
  {Horvati{\'{c}}}, \citenamefont {Berthier}, \citenamefont {Hardy},
  \citenamefont {Liang}, \citenamefont {Bonn},\ and\ \citenamefont
  {Julien}}]{Wu11_Nature}%
  \BibitemOpen
  \bibfield  {author} {\bibinfo {author} {\bibfnamefont {T.}~\bibnamefont
  {Wu}}, \bibinfo {author} {\bibfnamefont {H.}~\bibnamefont {Mayaffre}},
  \bibinfo {author} {\bibfnamefont {S.}~\bibnamefont {Kr{\"{a}}mer}}, \bibinfo
  {author} {\bibfnamefont {M.}~\bibnamefont {Horvati{\'{c}}}}, \bibinfo
  {author} {\bibfnamefont {C.}~\bibnamefont {Berthier}}, \bibinfo {author}
  {\bibfnamefont {W.~N.}\ \bibnamefont {Hardy}}, \bibinfo {author}
  {\bibfnamefont {R.}~\bibnamefont {Liang}}, \bibinfo {author} {\bibfnamefont
  {D.~A.}\ \bibnamefont {Bonn}}, \ and\ \bibinfo {author} {\bibfnamefont
  {M.-H.}\ \bibnamefont {Julien}},\ }\bibfield  {title} {\enquote {\bibinfo
  {title} {{Magnetic-field-induced charge-stripe order in the high-temperature
  superconductor YBa$_2$Cu$_3$O$_{\mathrm{\ensuremath{y}}}$}},}\ }\href
  {\doibase 10.1038/nature10345} {\bibfield  {journal} {\bibinfo  {journal}
  {Nature}\ }\textbf {\bibinfo {volume} {477}},\ \bibinfo {pages} {191--194}
  (\bibinfo {year} {2011})}\BibitemShut {NoStop}%
\bibitem [{\citenamefont {Gerber}\ \emph {et~al.}(2015)\citenamefont {Gerber},
  \citenamefont {Jang}, \citenamefont {Nojiri}, \citenamefont {Matsuzawa},
  \citenamefont {Yasumura}, \citenamefont {Bonn}, \citenamefont {Liang},
  \citenamefont {Hardy}, \citenamefont {Islam}, \citenamefont {Mehta},
  \citenamefont {Song}, \citenamefont {Sikorski}, \citenamefont {Stefanescu},
  \citenamefont {Feng}, \citenamefont {Kivelson}, \citenamefont {Devereaux},
  \citenamefont {Shen}, \citenamefont {Kao}, \citenamefont {Lee}, \citenamefont
  {Zhu},\ and\ \citenamefont {Lee}}]{Gerber15_Science}%
  \BibitemOpen
  \bibfield  {author} {\bibinfo {author} {\bibfnamefont {S.}~\bibnamefont
  {Gerber}}, \bibinfo {author} {\bibfnamefont {H.}~\bibnamefont {Jang}},
  \bibinfo {author} {\bibfnamefont {H.}~\bibnamefont {Nojiri}}, \bibinfo
  {author} {\bibfnamefont {S.}~\bibnamefont {Matsuzawa}}, \bibinfo {author}
  {\bibfnamefont {H.}~\bibnamefont {Yasumura}}, \bibinfo {author}
  {\bibfnamefont {D.~A.}\ \bibnamefont {Bonn}}, \bibinfo {author}
  {\bibfnamefont {R.}~\bibnamefont {Liang}}, \bibinfo {author} {\bibfnamefont
  {W.~N.}\ \bibnamefont {Hardy}}, \bibinfo {author} {\bibfnamefont
  {Z.}~\bibnamefont {Islam}}, \bibinfo {author} {\bibfnamefont
  {A.}~\bibnamefont {Mehta}}, \bibinfo {author} {\bibfnamefont
  {S.}~\bibnamefont {Song}}, \bibinfo {author} {\bibfnamefont {M.}~\bibnamefont
  {Sikorski}}, \bibinfo {author} {\bibfnamefont {D.}~\bibnamefont
  {Stefanescu}}, \bibinfo {author} {\bibfnamefont {Y.}~\bibnamefont {Feng}},
  \bibinfo {author} {\bibfnamefont {S.~A.}\ \bibnamefont {Kivelson}}, \bibinfo
  {author} {\bibfnamefont {T.~P.}\ \bibnamefont {Devereaux}}, \bibinfo {author}
  {\bibfnamefont {Z.-X.}\ \bibnamefont {Shen}}, \bibinfo {author}
  {\bibfnamefont {C.-C.}\ \bibnamefont {Kao}}, \bibinfo {author} {\bibfnamefont
  {W.-S.}\ \bibnamefont {Lee}}, \bibinfo {author} {\bibfnamefont
  {D.}~\bibnamefont {Zhu}}, \ and\ \bibinfo {author} {\bibfnamefont {J.-S.}\
  \bibnamefont {Lee}},\ }\bibfield  {title} {\enquote {\bibinfo {title}
  {{Three-dimensional charge density wave order in YBa$_2$Cu$_3$O$_{6.67}$ at
  high magnetic fields}},}\ }\href {\doibase 10.1126/science.aac6257}
  {\bibfield  {journal} {\bibinfo  {journal} {Science}\ }\textbf {\bibinfo
  {volume} {350}},\ \bibinfo {pages} {949--952} (\bibinfo {year}
  {2015})}\BibitemShut {NoStop}%
\bibitem [{\citenamefont {Chang}\ \emph {et~al.}(2016)\citenamefont {Chang},
  \citenamefont {Blackburn}, \citenamefont {Ivashko}, \citenamefont {Holmes},
  \citenamefont {Christensen}, \citenamefont {H{\"{u}}cker}, \citenamefont
  {Liang}, \citenamefont {Bonn}, \citenamefont {Hardy}, \citenamefont
  {R{\"{u}}tt}, \citenamefont {Zimmermann}, \citenamefont {Forgan},\ and\
  \citenamefont {Hayden}}]{Chang16_NatComm}%
  \BibitemOpen
  \bibfield  {author} {\bibinfo {author} {\bibfnamefont {J.}~\bibnamefont
  {Chang}}, \bibinfo {author} {\bibfnamefont {E.}~\bibnamefont {Blackburn}},
  \bibinfo {author} {\bibfnamefont {O.}~\bibnamefont {Ivashko}}, \bibinfo
  {author} {\bibfnamefont {A.~T.}\ \bibnamefont {Holmes}}, \bibinfo {author}
  {\bibfnamefont {N.~B.}\ \bibnamefont {Christensen}}, \bibinfo {author}
  {\bibfnamefont {M.}~\bibnamefont {H{\"{u}}cker}}, \bibinfo {author}
  {\bibfnamefont {R.}~\bibnamefont {Liang}}, \bibinfo {author} {\bibfnamefont
  {D.~A.}\ \bibnamefont {Bonn}}, \bibinfo {author} {\bibfnamefont {W.~N.}\
  \bibnamefont {Hardy}}, \bibinfo {author} {\bibfnamefont {U.}~\bibnamefont
  {R{\"{u}}tt}}, \bibinfo {author} {\bibfnamefont {M.~v.}\ \bibnamefont
  {Zimmermann}}, \bibinfo {author} {\bibfnamefont {E.~M.}\ \bibnamefont
  {Forgan}}, \ and\ \bibinfo {author} {\bibfnamefont {S.~M.}\ \bibnamefont
  {Hayden}},\ }\bibfield  {title} {\enquote {\bibinfo {title} {{Magnetic field
  controlled charge density wave coupling in underdoped
  YBa$_2$Cu$_3$O$_{6+x}$}},}\ }\href {\doibase 10.1038/ncomms11494} {\bibfield
  {journal} {\bibinfo  {journal} {Nat. Commun.}\ }\textbf {\bibinfo {volume}
  {7}},\ \bibinfo {pages} {11494} (\bibinfo {year} {2016})}\BibitemShut
  {NoStop}%
\bibitem [{\citenamefont {Kim}\ \emph {et~al.}(2018)\citenamefont {Kim},
  \citenamefont {Souliou}, \citenamefont {Barber}, \citenamefont
  {Lefran{\c{c}}ois}, \citenamefont {Minola}, \citenamefont {Tortora},
  \citenamefont {Heid}, \citenamefont {Nandi}, \citenamefont {Borzi},
  \citenamefont {Garbarino}, \citenamefont {Bosak}, \citenamefont {Porras},
  \citenamefont {Loew}, \citenamefont {K{\"{o}}nig}, \citenamefont {Moll},
  \citenamefont {Mackenzie}, \citenamefont {Keimer}, \citenamefont {Hicks},\
  and\ \citenamefont {Le~Tacon}}]{Kim18_Science}%
  \BibitemOpen
  \bibfield  {author} {\bibinfo {author} {\bibfnamefont {H.-H.}\ \bibnamefont
  {Kim}}, \bibinfo {author} {\bibfnamefont {S.~M.}\ \bibnamefont {Souliou}},
  \bibinfo {author} {\bibfnamefont {M.~E.}\ \bibnamefont {Barber}}, \bibinfo
  {author} {\bibfnamefont {E.}~\bibnamefont {Lefran{\c{c}}ois}}, \bibinfo
  {author} {\bibfnamefont {M.}~\bibnamefont {Minola}}, \bibinfo {author}
  {\bibfnamefont {M.}~\bibnamefont {Tortora}}, \bibinfo {author} {\bibfnamefont
  {R.}~\bibnamefont {Heid}}, \bibinfo {author} {\bibfnamefont {N.}~\bibnamefont
  {Nandi}}, \bibinfo {author} {\bibfnamefont {R.~A.}\ \bibnamefont {Borzi}},
  \bibinfo {author} {\bibfnamefont {G.}~\bibnamefont {Garbarino}}, \bibinfo
  {author} {\bibfnamefont {A.}~\bibnamefont {Bosak}}, \bibinfo {author}
  {\bibfnamefont {J.}~\bibnamefont {Porras}}, \bibinfo {author} {\bibfnamefont
  {T.}~\bibnamefont {Loew}}, \bibinfo {author} {\bibfnamefont {M.}~\bibnamefont
  {K{\"{o}}nig}}, \bibinfo {author} {\bibfnamefont {P.~J.~W.}\ \bibnamefont
  {Moll}}, \bibinfo {author} {\bibfnamefont {A.~P.}\ \bibnamefont {Mackenzie}},
  \bibinfo {author} {\bibfnamefont {B.}~\bibnamefont {Keimer}}, \bibinfo
  {author} {\bibfnamefont {C.~W.}\ \bibnamefont {Hicks}}, \ and\ \bibinfo
  {author} {\bibfnamefont {M.}~\bibnamefont {Le~Tacon}},\ }\bibfield  {title}
  {\enquote {\bibinfo {title} {{Uniaxial pressure control of competing orders
  in a high-temperature superconductor}},}\ }\href {\doibase
  10.1126/science.aat4708} {\bibfield  {journal} {\bibinfo  {journal}
  {Science}\ }\textbf {\bibinfo {volume} {362}},\ \bibinfo {pages} {1040--1044}
  (\bibinfo {year} {2018})}\BibitemShut {NoStop}%
\bibitem [{\citenamefont {Castro~Neto}(2001)}]{CastroNeto01_PRB}%
  \BibitemOpen
  \bibfield  {author} {\bibinfo {author} {\bibfnamefont {A.~H.}\ \bibnamefont
  {Castro~Neto}},\ }\bibfield  {title} {\enquote {\bibinfo {title} {{Stripes,
  vibrations, and superconductivity}},}\ }\href {\doibase
  10.1103/PhysRevB.64.104509} {\bibfield  {journal} {\bibinfo  {journal} {Phys.
  Rev. B}\ }\textbf {\bibinfo {volume} {64}},\ \bibinfo {pages} {104509}
  (\bibinfo {year} {2001})}\BibitemShut {NoStop}%
\bibitem [{\citenamefont {Ghiringhelli}\ \emph {et~al.}(2012)\citenamefont
  {Ghiringhelli}, \citenamefont {Le~Tacon}, \citenamefont {Minola},
  \citenamefont {Blanco-Canosa}, \citenamefont {Mazzoli}, \citenamefont
  {Brookes}, \citenamefont {De~Luca}, \citenamefont {Frano}, \citenamefont
  {Hawthorn}, \citenamefont {He}, \citenamefont {Loew}, \citenamefont {Sala},
  \citenamefont {Peets}, \citenamefont {Salluzzo}, \citenamefont {Schierle},
  \citenamefont {Sutarto}, \citenamefont {Sawatzky}, \citenamefont {Weschke},
  \citenamefont {Keimer},\ and\ \citenamefont
  {Braicovich}}]{Ghiringhelli12_Science}%
  \BibitemOpen
  \bibfield  {author} {\bibinfo {author} {\bibfnamefont {G.}~\bibnamefont
  {Ghiringhelli}}, \bibinfo {author} {\bibfnamefont {M.}~\bibnamefont
  {Le~Tacon}}, \bibinfo {author} {\bibfnamefont {M.}~\bibnamefont {Minola}},
  \bibinfo {author} {\bibfnamefont {S.}~\bibnamefont {Blanco-Canosa}}, \bibinfo
  {author} {\bibfnamefont {C.}~\bibnamefont {Mazzoli}}, \bibinfo {author}
  {\bibfnamefont {N.~B.}\ \bibnamefont {Brookes}}, \bibinfo {author}
  {\bibfnamefont {G.~M.}\ \bibnamefont {De~Luca}}, \bibinfo {author}
  {\bibfnamefont {A.}~\bibnamefont {Frano}}, \bibinfo {author} {\bibfnamefont
  {D.~G.}\ \bibnamefont {Hawthorn}}, \bibinfo {author} {\bibfnamefont
  {F.}~\bibnamefont {He}}, \bibinfo {author} {\bibfnamefont {T.}~\bibnamefont
  {Loew}}, \bibinfo {author} {\bibfnamefont {M.~M.}\ \bibnamefont {Sala}},
  \bibinfo {author} {\bibfnamefont {D.~C.}\ \bibnamefont {Peets}}, \bibinfo
  {author} {\bibfnamefont {M.}~\bibnamefont {Salluzzo}}, \bibinfo {author}
  {\bibfnamefont {E.}~\bibnamefont {Schierle}}, \bibinfo {author}
  {\bibfnamefont {R.}~\bibnamefont {Sutarto}}, \bibinfo {author} {\bibfnamefont
  {G.~A.}\ \bibnamefont {Sawatzky}}, \bibinfo {author} {\bibfnamefont
  {E.}~\bibnamefont {Weschke}}, \bibinfo {author} {\bibfnamefont
  {B.}~\bibnamefont {Keimer}}, \ and\ \bibinfo {author} {\bibfnamefont
  {L.}~\bibnamefont {Braicovich}},\ }\bibfield  {title} {\enquote {\bibinfo
  {title} {{Long-Range Incommensurate Charge Fluctuations in
  (Y,Nd)Ba$_2$Cu$_3$O$_{6+x}$}},}\ }\href {\doibase 10.1126/science.1223532}
  {\bibfield  {journal} {\bibinfo  {journal} {Science}\ }\textbf {\bibinfo
  {volume} {337}},\ \bibinfo {pages} {821--825} (\bibinfo {year}
  {2012})}\BibitemShut {NoStop}%
\bibitem [{\citenamefont {Croft}\ \emph {et~al.}(2014)\citenamefont {Croft},
  \citenamefont {Lester}, \citenamefont {Senn}, \citenamefont {Bombardi},\ and\
  \citenamefont {Hayden}}]{Croft14_PRB}%
  \BibitemOpen
  \bibfield  {author} {\bibinfo {author} {\bibfnamefont {T.~P.}\ \bibnamefont
  {Croft}}, \bibinfo {author} {\bibfnamefont {C.}~\bibnamefont {Lester}},
  \bibinfo {author} {\bibfnamefont {M.~S.}\ \bibnamefont {Senn}}, \bibinfo
  {author} {\bibfnamefont {A.}~\bibnamefont {Bombardi}}, \ and\ \bibinfo
  {author} {\bibfnamefont {S.~M.}\ \bibnamefont {Hayden}},\ }\bibfield  {title}
  {\enquote {\bibinfo {title} {{Charge density wave fluctuations in
  ${\text{La}}_{2\ensuremath{-}x}$${\text{Sr}}_{x}$${\text{CuO}}_{4}$ and their
  competition with superconductivity}},}\ }\href {\doibase
  10.1103/PhysRevB.89.224513} {\bibfield  {journal} {\bibinfo  {journal} {Phys.
  Rev. B}\ }\textbf {\bibinfo {volume} {89}},\ \bibinfo {pages} {224513}
  (\bibinfo {year} {2014})}\BibitemShut {NoStop}%
\bibitem [{\citenamefont {Chang}\ \emph {et~al.}(2012)\citenamefont {Chang},
  \citenamefont {Blackburn}, \citenamefont {Holmes}, \citenamefont
  {Christensen}, \citenamefont {Larsen}, \citenamefont {Mesot}, \citenamefont
  {Liang}, \citenamefont {Bonn}, \citenamefont {Hardy}, \citenamefont
  {Watenphul}, \citenamefont {Zimmermann}, \citenamefont {Forgan},\ and\
  \citenamefont {Hayden}}]{Chang12_NatPhys}%
  \BibitemOpen
  \bibfield  {author} {\bibinfo {author} {\bibfnamefont {J.}~\bibnamefont
  {Chang}}, \bibinfo {author} {\bibfnamefont {E.}~\bibnamefont {Blackburn}},
  \bibinfo {author} {\bibfnamefont {A.~T.}\ \bibnamefont {Holmes}}, \bibinfo
  {author} {\bibfnamefont {N.~B.}\ \bibnamefont {Christensen}}, \bibinfo
  {author} {\bibfnamefont {J.}~\bibnamefont {Larsen}}, \bibinfo {author}
  {\bibfnamefont {J.}~\bibnamefont {Mesot}}, \bibinfo {author} {\bibfnamefont
  {R.}~\bibnamefont {Liang}}, \bibinfo {author} {\bibfnamefont {D.~A.}\
  \bibnamefont {Bonn}}, \bibinfo {author} {\bibfnamefont {W.~N.}\ \bibnamefont
  {Hardy}}, \bibinfo {author} {\bibfnamefont {A.}~\bibnamefont {Watenphul}},
  \bibinfo {author} {\bibfnamefont {M.~v.}\ \bibnamefont {Zimmermann}},
  \bibinfo {author} {\bibfnamefont {E.~M.}\ \bibnamefont {Forgan}}, \ and\
  \bibinfo {author} {\bibfnamefont {S.~M.}\ \bibnamefont {Hayden}},\ }\bibfield
   {title} {\enquote {\bibinfo {title} {{Direct observation of competition
  between superconductivity and charge density wave order in
  YBa$_2$Cu$_3$O$_{6.67}$}},}\ }\href {\doibase 10.1038/nphys2456} {\bibfield
  {journal} {\bibinfo  {journal} {Nat. Phys.}\ }\textbf {\bibinfo {volume}
  {8}},\ \bibinfo {pages} {871--876} (\bibinfo {year} {2012})}\BibitemShut
  {NoStop}%
\bibitem [{\citenamefont {Liang}\ \emph {et~al.}(2006)\citenamefont {Liang},
  \citenamefont {Bonn},\ and\ \citenamefont {Hardy}}]{Liang06_PRB}%
  \BibitemOpen
  \bibfield  {author} {\bibinfo {author} {\bibfnamefont {R.}~\bibnamefont
  {Liang}}, \bibinfo {author} {\bibfnamefont {D.~A.}\ \bibnamefont {Bonn}}, \
  and\ \bibinfo {author} {\bibfnamefont {W.~N.}\ \bibnamefont {Hardy}},\
  }\bibfield  {title} {\enquote {\bibinfo {title} {{Evaluation of
  ${\mathrm{CuO}}_{2}$ plane hole doping in
  ${\mathrm{YBa}}_{2}{\mathrm{Cu}}_{3}{\mathrm{O}}_{6+x}$ single crystals}},}\
  }\href {\doibase 10.1103/PhysRevB.73.180505} {\bibfield  {journal} {\bibinfo
  {journal} {Phys. Rev. B}\ }\textbf {\bibinfo {volume} {73}},\ \bibinfo
  {pages} {180505(R)} (\bibinfo {year} {2006})}\BibitemShut {NoStop}%
\bibitem [{\citenamefont {Cyr-Choini{\`{e}}re}\ \emph
  {et~al.}(2018)\citenamefont {Cyr-Choini{\`{e}}re}, \citenamefont {LeBoeuf},
  \citenamefont {Badoux}, \citenamefont {Dufour-Beaus{\'{e}}jour},
  \citenamefont {Bonn}, \citenamefont {Hardy}, \citenamefont {Liang},
  \citenamefont {Graf}, \citenamefont {Doiron-Leyraud},\ and\ \citenamefont
  {Taillefer}}]{CyrChoiniere18_PRB}%
  \BibitemOpen
  \bibfield  {author} {\bibinfo {author} {\bibfnamefont {O.}~\bibnamefont
  {Cyr-Choini{\`{e}}re}}, \bibinfo {author} {\bibfnamefont {D.}~\bibnamefont
  {LeBoeuf}}, \bibinfo {author} {\bibfnamefont {S.}~\bibnamefont {Badoux}},
  \bibinfo {author} {\bibfnamefont {S.}~\bibnamefont
  {Dufour-Beaus{\'{e}}jour}}, \bibinfo {author} {\bibfnamefont {D.~A.}\
  \bibnamefont {Bonn}}, \bibinfo {author} {\bibfnamefont {W.~N.}\ \bibnamefont
  {Hardy}}, \bibinfo {author} {\bibfnamefont {R.}~\bibnamefont {Liang}},
  \bibinfo {author} {\bibfnamefont {D.}~\bibnamefont {Graf}}, \bibinfo {author}
  {\bibfnamefont {N.}~\bibnamefont {Doiron-Leyraud}}, \ and\ \bibinfo {author}
  {\bibfnamefont {L.}~\bibnamefont {Taillefer}},\ }\bibfield  {title} {\enquote
  {\bibinfo {title} {{Sensitivity of ${T}_{\mathrm{c}}$ to pressure and
  magnetic field in the cuprate superconductor
  ${\mathrm{YBa}}_{2}{\mathrm{Cu}}_{3}{\mathrm{O}}_{y}$: Evidence of
  charge-order suppression by pressure}},}\ }\href {\doibase
  10.1103/PhysRevB.98.064513} {\bibfield  {journal} {\bibinfo  {journal} {Phys.
  Rev. B}\ }\textbf {\bibinfo {volume} {98}},\ \bibinfo {pages} {064513}
  (\bibinfo {year} {2018})}\BibitemShut {NoStop}%
\bibitem [{\citenamefont {Souliou}\ \emph {et~al.}(2018)\citenamefont
  {Souliou}, \citenamefont {Gretarsson}, \citenamefont {Garbarino},
  \citenamefont {Bosak}, \citenamefont {Porras}, \citenamefont {Loew},
  \citenamefont {Keimer},\ and\ \citenamefont {Le~Tacon}}]{Souliou18_PRB}%
  \BibitemOpen
  \bibfield  {author} {\bibinfo {author} {\bibfnamefont {S.~M.}\ \bibnamefont
  {Souliou}}, \bibinfo {author} {\bibfnamefont {H.}~\bibnamefont {Gretarsson}},
  \bibinfo {author} {\bibfnamefont {G.}~\bibnamefont {Garbarino}}, \bibinfo
  {author} {\bibfnamefont {A.}~\bibnamefont {Bosak}}, \bibinfo {author}
  {\bibfnamefont {J.}~\bibnamefont {Porras}}, \bibinfo {author} {\bibfnamefont
  {T.}~\bibnamefont {Loew}}, \bibinfo {author} {\bibfnamefont {B.}~\bibnamefont
  {Keimer}}, \ and\ \bibinfo {author} {\bibfnamefont {M.}~\bibnamefont
  {Le~Tacon}},\ }\bibfield  {title} {\enquote {\bibinfo {title} {{Rapid
  suppression of the charge density wave in
  ${\mathrm{YBa}}_{2}{\mathrm{Cu}}_{3}{\mathrm{O}}_{6.6}$ under hydrostatic
  pressure}},}\ }\href {\doibase 10.1103/PhysRevB.97.020503} {\bibfield
  {journal} {\bibinfo  {journal} {Phys. Rev. B}\ }\textbf {\bibinfo {volume}
  {97}},\ \bibinfo {pages} {020503(R)} (\bibinfo {year} {2018})}\BibitemShut
  {NoStop}%
\bibitem [{\citenamefont {Grissonnanche}\ \emph {et~al.}(2014)\citenamefont
  {Grissonnanche}, \citenamefont {Cyr-Choini{\`{e}}re}, \citenamefont
  {Lalibert{\'{e}}}, \citenamefont {Ren{\'{e}}~de Cotret}, \citenamefont
  {Juneau-Fecteau}, \citenamefont {Dufour-Beaus{\'{e}}jour}, \citenamefont
  {Delage}, \citenamefont {LeBoeuf}, \citenamefont {Chang}, \citenamefont
  {Ramshaw}, \citenamefont {Bonn}, \citenamefont {Hardy}, \citenamefont
  {Liang}, \citenamefont {Adachi}, \citenamefont {Hussey}, \citenamefont
  {Vignolle}, \citenamefont {Proust}, \citenamefont {Kr{\"{a}}mer},
  \citenamefont {Park}, \citenamefont {Graf}, \citenamefont {Doiron-Leyraud},\
  and\ \citenamefont {Taillefer}}]{Grissonnanche14_NatComm}%
  \BibitemOpen
  \bibfield  {author} {\bibinfo {author} {\bibfnamefont {G.}~\bibnamefont
  {Grissonnanche}}, \bibinfo {author} {\bibfnamefont {O.}~\bibnamefont
  {Cyr-Choini{\`{e}}re}}, \bibinfo {author} {\bibfnamefont {F.}~\bibnamefont
  {Lalibert{\'{e}}}}, \bibinfo {author} {\bibfnamefont {S.}~\bibnamefont
  {Ren{\'{e}}~de Cotret}}, \bibinfo {author} {\bibfnamefont {A.}~\bibnamefont
  {Juneau-Fecteau}}, \bibinfo {author} {\bibfnamefont {S.}~\bibnamefont
  {Dufour-Beaus{\'{e}}jour}}, \bibinfo {author} {\bibfnamefont {M.-{\`{E}}.}\
  \bibnamefont {Delage}}, \bibinfo {author} {\bibfnamefont {D.}~\bibnamefont
  {LeBoeuf}}, \bibinfo {author} {\bibfnamefont {J.}~\bibnamefont {Chang}},
  \bibinfo {author} {\bibfnamefont {B.~J.}\ \bibnamefont {Ramshaw}}, \bibinfo
  {author} {\bibfnamefont {D.~A.}\ \bibnamefont {Bonn}}, \bibinfo {author}
  {\bibfnamefont {W.~N.}\ \bibnamefont {Hardy}}, \bibinfo {author}
  {\bibfnamefont {R.}~\bibnamefont {Liang}}, \bibinfo {author} {\bibfnamefont
  {S.}~\bibnamefont {Adachi}}, \bibinfo {author} {\bibfnamefont {N.~E.}\
  \bibnamefont {Hussey}}, \bibinfo {author} {\bibfnamefont {B.}~\bibnamefont
  {Vignolle}}, \bibinfo {author} {\bibfnamefont {M.}~\bibnamefont {Proust},
  \bibfnamefont {C.~amd~Sutherland}}, \bibinfo {author} {\bibfnamefont
  {S.}~\bibnamefont {Kr{\"{a}}mer}}, \bibinfo {author} {\bibfnamefont {J.-H.}\
  \bibnamefont {Park}}, \bibinfo {author} {\bibfnamefont {D.}~\bibnamefont
  {Graf}}, \bibinfo {author} {\bibfnamefont {N.}~\bibnamefont
  {Doiron-Leyraud}}, \ and\ \bibinfo {author} {\bibfnamefont {L.}~\bibnamefont
  {Taillefer}},\ }\bibfield  {title} {\enquote {\bibinfo {title} {{Direct
  measurement of the upper critical field in cuprate superconductors}},}\
  }\href {\doibase 10.1038/ncomms4280} {\bibfield  {journal} {\bibinfo
  {journal} {Nat. Commun.}\ }\textbf {\bibinfo {volume} {5}},\ \bibinfo {pages}
  {3280} (\bibinfo {year} {2014})}\BibitemShut {NoStop}%
\bibitem [{\citenamefont {Ka{\v{c}}mar{\v{c}}{\'{\i}}k}\ \emph
  {et~al.}(2018)\citenamefont {Ka{\v{c}}mar{\v{c}}{\'{\i}}k}, \citenamefont
  {Vinograd}, \citenamefont {Michon}, \citenamefont {Rydh}, \citenamefont
  {Demuer}, \citenamefont {Zhou}, \citenamefont {Mayaffre}, \citenamefont
  {Liang}, \citenamefont {Hardy}, \citenamefont {Bonn}, \citenamefont
  {Doiron-Leyraud}, \citenamefont {Taillefer}, \citenamefont {Julien},
  \citenamefont {Marcenat},\ and\ \citenamefont {Klein}}]{Kacmarcik18_PRL}%
  \BibitemOpen
  \bibfield  {author} {\bibinfo {author} {\bibfnamefont {J.}~\bibnamefont
  {Ka{\v{c}}mar{\v{c}}{\'{\i}}k}}, \bibinfo {author} {\bibfnamefont
  {I.}~\bibnamefont {Vinograd}}, \bibinfo {author} {\bibfnamefont
  {B.}~\bibnamefont {Michon}}, \bibinfo {author} {\bibfnamefont
  {A.}~\bibnamefont {Rydh}}, \bibinfo {author} {\bibfnamefont {A.}~\bibnamefont
  {Demuer}}, \bibinfo {author} {\bibfnamefont {R.}~\bibnamefont {Zhou}},
  \bibinfo {author} {\bibfnamefont {H.}~\bibnamefont {Mayaffre}}, \bibinfo
  {author} {\bibfnamefont {R.}~\bibnamefont {Liang}}, \bibinfo {author}
  {\bibfnamefont {W.~N.}\ \bibnamefont {Hardy}}, \bibinfo {author}
  {\bibfnamefont {D.~A.}\ \bibnamefont {Bonn}}, \bibinfo {author}
  {\bibfnamefont {N.}~\bibnamefont {Doiron-Leyraud}}, \bibinfo {author}
  {\bibfnamefont {L.}~\bibnamefont {Taillefer}}, \bibinfo {author}
  {\bibfnamefont {M.-H.}\ \bibnamefont {Julien}}, \bibinfo {author}
  {\bibfnamefont {C.}~\bibnamefont {Marcenat}}, \ and\ \bibinfo {author}
  {\bibfnamefont {T.}~\bibnamefont {Klein}},\ }\bibfield  {title} {\enquote
  {\bibinfo {title} {{Unusual Interplay between Superconductivity and
  Field-Induced Charge Order in
  ${\mathrm{YBa}}_{2}{\mathrm{Cu}}_{3}{\mathrm{O}}_{y}$}},}\ }\href {\doibase
  10.1103/PhysRevLett.121.167002} {\bibfield  {journal} {\bibinfo  {journal}
  {Phys. Rev. Lett.}\ }\textbf {\bibinfo {volume} {121}},\ \bibinfo {pages}
  {167002} (\bibinfo {year} {2018})}\BibitemShut {NoStop}%
\bibitem [{\citenamefont {Choi}\ \emph {et~al.}(2020)\citenamefont {Choi},
  \citenamefont {Ivashko}, \citenamefont {Blackburn}, \citenamefont {Liang},
  \citenamefont {Bonn}, \citenamefont {Hardy}, \citenamefont {Holmes},
  \citenamefont {Christensen}, \citenamefont {H{\"{u}}cker}, \citenamefont
  {Gerber}, \citenamefont {Gutowski}, \citenamefont {R{\"{u}}tt}, \citenamefont
  {Zimmermann}, \citenamefont {Forgan}, \citenamefont {Hayden},\ and\
  \citenamefont {Chang}}]{Choi20_NatComm}%
  \BibitemOpen
  \bibfield  {author} {\bibinfo {author} {\bibfnamefont {J.}~\bibnamefont
  {Choi}}, \bibinfo {author} {\bibfnamefont {O.}~\bibnamefont {Ivashko}},
  \bibinfo {author} {\bibfnamefont {E.}~\bibnamefont {Blackburn}}, \bibinfo
  {author} {\bibfnamefont {R.}~\bibnamefont {Liang}}, \bibinfo {author}
  {\bibfnamefont {D.~A.}\ \bibnamefont {Bonn}}, \bibinfo {author}
  {\bibfnamefont {W.~N.}\ \bibnamefont {Hardy}}, \bibinfo {author}
  {\bibfnamefont {A.~T.}\ \bibnamefont {Holmes}}, \bibinfo {author}
  {\bibfnamefont {N.~B.}\ \bibnamefont {Christensen}}, \bibinfo {author}
  {\bibfnamefont {M.}~\bibnamefont {H{\"{u}}cker}}, \bibinfo {author}
  {\bibfnamefont {S.}~\bibnamefont {Gerber}}, \bibinfo {author} {\bibfnamefont
  {O.}~\bibnamefont {Gutowski}}, \bibinfo {author} {\bibfnamefont
  {U.}~\bibnamefont {R{\"{u}}tt}}, \bibinfo {author} {\bibfnamefont {M.~v.}\
  \bibnamefont {Zimmermann}}, \bibinfo {author} {\bibfnamefont {E.~M.}\
  \bibnamefont {Forgan}}, \bibinfo {author} {\bibfnamefont {S.~M.}\
  \bibnamefont {Hayden}}, \ and\ \bibinfo {author} {\bibfnamefont
  {J.}~\bibnamefont {Chang}},\ }\bibfield  {title} {\enquote {\bibinfo {title}
  {{Spatially inhomogeneous competition between superconductivity and the
  charge density wave in YBa$_2$Cu$_3$O$_{6.67}$}},}\ }\href {\doibase
  10.1038/s41467-020-14536-1} {\bibfield  {journal} {\bibinfo  {journal} {Nat.
  Commun.}\ }\textbf {\bibinfo {volume} {11}},\ \bibinfo {pages} {990}
  (\bibinfo {year} {2020})}\BibitemShut {NoStop}%
\bibitem [{\citenamefont {Okazaki}\ \emph {et~al.}(2007)\citenamefont
  {Okazaki}, \citenamefont {Shishido}, \citenamefont {Shibauchi}, \citenamefont
  {Konczykowski}, \citenamefont {Buzdin},\ and\ \citenamefont
  {Matsuda}}]{Okazaki07_PRB}%
  \BibitemOpen
  \bibfield  {author} {\bibinfo {author} {\bibfnamefont {R.}~\bibnamefont
  {Okazaki}}, \bibinfo {author} {\bibfnamefont {H.}~\bibnamefont {Shishido}},
  \bibinfo {author} {\bibfnamefont {T.}~\bibnamefont {Shibauchi}}, \bibinfo
  {author} {\bibfnamefont {M.}~\bibnamefont {Konczykowski}}, \bibinfo {author}
  {\bibfnamefont {A.}~\bibnamefont {Buzdin}}, \ and\ \bibinfo {author}
  {\bibfnamefont {Y.}~\bibnamefont {Matsuda}},\ }\bibfield  {title} {\enquote
  {\bibinfo {title} {{High-field superconducting transition of
  $\mathrm{Ce}\mathrm{Co}{\mathrm{In}}_{5}$ studied by local magnetic induction
  measurements}},}\ }\href {\doibase 10.1103/PhysRevB.76.224529} {\bibfield
  {journal} {\bibinfo  {journal} {Phys. Rev. B}\ }\textbf {\bibinfo {volume}
  {76}},\ \bibinfo {pages} {224529} (\bibinfo {year} {2007})}\BibitemShut
  {NoStop}%
\bibitem [{\citenamefont {Gomes}\ \emph {et~al.}(2007)\citenamefont {Gomes},
  \citenamefont {Pasupathy}, \citenamefont {Pushp}, \citenamefont {Ono},
  \citenamefont {Ando},\ and\ \citenamefont {Yazdani}}]{Gomes07_Nature}%
  \BibitemOpen
  \bibfield  {author} {\bibinfo {author} {\bibfnamefont {K.~K.}\ \bibnamefont
  {Gomes}}, \bibinfo {author} {\bibfnamefont {A.~N.}\ \bibnamefont
  {Pasupathy}}, \bibinfo {author} {\bibfnamefont {A.}~\bibnamefont {Pushp}},
  \bibinfo {author} {\bibfnamefont {S.}~\bibnamefont {Ono}}, \bibinfo {author}
  {\bibfnamefont {Y.}~\bibnamefont {Ando}}, \ and\ \bibinfo {author}
  {\bibfnamefont {A.}~\bibnamefont {Yazdani}},\ }\bibfield  {title} {\enquote
  {\bibinfo {title} {{Visualizing pair formation on the atomic scale in the
  high-$T_{\mathrm{c}}$ superconductor Bi$_2$Sr$_2$CaCu$_2$O$_{8+\delta}$}},}\
  }\href {\doibase 10.1038/nature05881} {\bibfield  {journal} {\bibinfo
  {journal} {Nature}\ }\textbf {\bibinfo {volume} {447}},\ \bibinfo {pages}
  {569--572} (\bibinfo {year} {2007})}\BibitemShut {NoStop}%
\bibitem [{\citenamefont {Kittaka}\ \emph {et~al.}(2010)\citenamefont
  {Kittaka}, \citenamefont {Taniguchi}, \citenamefont {Yonezawa}, \citenamefont
  {Yaguchi},\ and\ \citenamefont {Maeno}}]{Kittaka10_PRB}%
  \BibitemOpen
  \bibfield  {author} {\bibinfo {author} {\bibfnamefont {S.}~\bibnamefont
  {Kittaka}}, \bibinfo {author} {\bibfnamefont {H.}~\bibnamefont {Taniguchi}},
  \bibinfo {author} {\bibfnamefont {S.}~\bibnamefont {Yonezawa}}, \bibinfo
  {author} {\bibfnamefont {H.}~\bibnamefont {Yaguchi}}, \ and\ \bibinfo
  {author} {\bibfnamefont {Y.}~\bibnamefont {Maeno}},\ }\bibfield  {title}
  {\enquote {\bibinfo {title} {{Higher-${T}_{c}$ superconducting phase in
  ${\text{Sr}}_{2}{\text{RuO}}_{4}$ induced by uniaxial pressure}},}\ }\href
  {\doibase 10.1103/PhysRevB.81.180510} {\bibfield  {journal} {\bibinfo
  {journal} {Phys. Rev. B}\ }\textbf {\bibinfo {volume} {81}},\ \bibinfo
  {pages} {180510(R)} (\bibinfo {year} {2010})}\BibitemShut {NoStop}%
\bibitem [{\citenamefont {Metzger}\ \emph {et~al.}(1993)\citenamefont
  {Metzger}, \citenamefont {Weber}, \citenamefont {Fietz}, \citenamefont
  {Grube}, \citenamefont {Ludwig}, \citenamefont {Wolf},\ and\ \citenamefont
  {W{\"{u}}hl}}]{Metzger93_PhysicaC}%
  \BibitemOpen
  \bibfield  {author} {\bibinfo {author} {\bibfnamefont {J.}~\bibnamefont
  {Metzger}}, \bibinfo {author} {\bibfnamefont {T.}~\bibnamefont {Weber}},
  \bibinfo {author} {\bibfnamefont {W.}~\bibnamefont {Fietz}}, \bibinfo
  {author} {\bibfnamefont {K.}~\bibnamefont {Grube}}, \bibinfo {author}
  {\bibfnamefont {H.}~\bibnamefont {Ludwig}}, \bibinfo {author} {\bibfnamefont
  {T.}~\bibnamefont {Wolf}}, \ and\ \bibinfo {author} {\bibfnamefont
  {H.}~\bibnamefont {W{\"{u}}hl}},\ }\bibfield  {title} {\enquote {\bibinfo
  {title} {{Separation of the intrinsic pressure effect on $T_{\mathrm{c}}$ of
  YBa$_2$Cu$_3$O$_{6.7}$ from a $T_{\mathrm{c}}$ enhancement caused by
  pressure-induced oxygen ordering}},}\ }\href {\doibase
  10.1016/0921-4534(93)90840-M} {\bibfield  {journal} {\bibinfo  {journal}
  {Physica C}\ }\textbf {\bibinfo {volume} {214}},\ \bibinfo {pages} {371--376}
  (\bibinfo {year} {1993})}\BibitemShut {NoStop}%
\bibitem [{\citenamefont {Fietz}\ \emph {et~al.}(1994)\citenamefont {Fietz},
  \citenamefont {Metzger}, \citenamefont {Weber}, \citenamefont {Grube},\ and\
  \citenamefont {Ludwig}}]{Fietz94_AIPConfProc}%
  \BibitemOpen
  \bibfield  {author} {\bibinfo {author} {\bibfnamefont {W.~H.}\ \bibnamefont
  {Fietz}}, \bibinfo {author} {\bibfnamefont {J.}~\bibnamefont {Metzger}},
  \bibinfo {author} {\bibfnamefont {T.}~\bibnamefont {Weber}}, \bibinfo
  {author} {\bibfnamefont {K.}~\bibnamefont {Grube}}, \ and\ \bibinfo {author}
  {\bibfnamefont {H.~A.}\ \bibnamefont {Ludwig}},\ }\bibfield  {title}
  {\enquote {\bibinfo {title} {{Dependence of the intrinsic
  $dT_{\mathrm{c}}/dp$ of YBa$_2$Cu$_3$O$_x$ on the oxygen content and the
  additive $T_{\mathrm{c}}$ increase by pressure‐induced oxygen ordering}},}\
  }\href {\doibase 10.1063/1.46430} {\bibfield  {journal} {\bibinfo  {journal}
  {AIP Conf. Proc.}\ }\textbf {\bibinfo {volume} {309}},\ \bibinfo {pages}
  {703--706} (\bibinfo {year} {1994})}\BibitemShut {NoStop}%
\bibitem [{\citenamefont {Kr\"{u}ger}\ \emph {et~al.}(1997)\citenamefont
  {Kr\"{u}ger}, \citenamefont {Conder}, \citenamefont {Schwer},\ and\
  \citenamefont {Kaldis}}]{Krueger97_JSSC}%
  \BibitemOpen
  \bibfield  {author} {\bibinfo {author} {\bibfnamefont {C.}~\bibnamefont
  {Kr\"{u}ger}}, \bibinfo {author} {\bibfnamefont {K.}~\bibnamefont {Conder}},
  \bibinfo {author} {\bibfnamefont {H.}~\bibnamefont {Schwer}}, \ and\ \bibinfo
  {author} {\bibfnamefont {E.}~\bibnamefont {Kaldis}},\ }\bibfield  {title}
  {\enquote {\bibinfo {title} {{The Dependence of the Lattice Parameters on
  Oxygen Content in Orthorhombic
  ${\mathrm{YBa}}_{2}$${\mathrm{Cu}}_{3}$${\mathrm{O}}_{6\mathrm{\ensuremath{+}}\mathrm{\ensuremath{x}}}$:
  A High Precision Reinvestigation of Near Equilibrium Samples}},}\ }\href
  {\doibase 10.1006/jssc.1997.7579} {\bibfield  {journal} {\bibinfo  {journal}
  {J. Solid State Chem.}\ }\textbf {\bibinfo {volume} {134}},\ \bibinfo {pages}
  {356--361} (\bibinfo {year} {1997})}\BibitemShut {NoStop}%
\bibitem [{\citenamefont {Meingast}\ \emph {et~al.}(1991)\citenamefont
  {Meingast}, \citenamefont {Kraut}, \citenamefont {Wolf}, \citenamefont
  {W{\"{u}}hl}, \citenamefont {Erb},\ and\ \citenamefont
  {M{\"{u}}ller-Vogt}}]{Meingast91_PRL}%
  \BibitemOpen
  \bibfield  {author} {\bibinfo {author} {\bibfnamefont {C.}~\bibnamefont
  {Meingast}}, \bibinfo {author} {\bibfnamefont {O.}~\bibnamefont {Kraut}},
  \bibinfo {author} {\bibfnamefont {T.}~\bibnamefont {Wolf}}, \bibinfo {author}
  {\bibfnamefont {H.}~\bibnamefont {W{\"{u}}hl}}, \bibinfo {author}
  {\bibfnamefont {A.}~\bibnamefont {Erb}}, \ and\ \bibinfo {author}
  {\bibfnamefont {G.}~\bibnamefont {M{\"{u}}ller-Vogt}},\ }\bibfield  {title}
  {\enquote {\bibinfo {title} {{Large $a$-$b$ anisotropy of the expansivity
  anomaly at ${\mathit{T}}_{\mathrm{c}}$ in untwinned
  ${\mathrm{YBa}}_{2}$${\mathrm{Cu}}_{3}$${\mathrm{O}}_{7\mathrm{\ensuremath{-}}\mathrm{\ensuremath{\delta}}}$}},}\
  }\href {\doibase 10.1103/PhysRevLett.67.1634} {\bibfield  {journal} {\bibinfo
   {journal} {Phys. Rev. Lett.}\ }\textbf {\bibinfo {volume} {67}},\ \bibinfo
  {pages} {1634--1637} (\bibinfo {year} {1991})}\BibitemShut {NoStop}%
\bibitem [{\citenamefont {Welp}\ \emph {et~al.}(1994)\citenamefont {Welp},
  \citenamefont {Grimsditch}, \citenamefont {Fleshler}, \citenamefont
  {Nessler}, \citenamefont {Veal},\ and\ \citenamefont
  {Crabtree}}]{Welp94_JSupercond}%
  \BibitemOpen
  \bibfield  {author} {\bibinfo {author} {\bibfnamefont {U.}~\bibnamefont
  {Welp}}, \bibinfo {author} {\bibfnamefont {M.}~\bibnamefont {Grimsditch}},
  \bibinfo {author} {\bibfnamefont {S.}~\bibnamefont {Fleshler}}, \bibinfo
  {author} {\bibfnamefont {W.}~\bibnamefont {Nessler}}, \bibinfo {author}
  {\bibfnamefont {B.}~\bibnamefont {Veal}}, \ and\ \bibinfo {author}
  {\bibfnamefont {G.~W.}\ \bibnamefont {Crabtree}},\ }\bibfield  {title}
  {\enquote {\bibinfo {title} {{Anisotropic uniaxial pressure effects in
  YBa$_2$Cu$_3$O$_{7-\delta}$}},}\ }\href {\doibase 10.1007/BF00730387}
  {\bibfield  {journal} {\bibinfo  {journal} {J. Supercond.}\ }\textbf
  {\bibinfo {volume} {7}},\ \bibinfo {pages} {159--164} (\bibinfo {year}
  {1994})}\BibitemShut {NoStop}%
\bibitem [{\citenamefont {Coneri}\ \emph {et~al.}(2010)\citenamefont {Coneri},
  \citenamefont {Sanna}, \citenamefont {Zheng}, \citenamefont {Lord},\ and\
  \citenamefont {De~Renzi}}]{Coneri10_PRB}%
  \BibitemOpen
  \bibfield  {author} {\bibinfo {author} {\bibfnamefont {F.}~\bibnamefont
  {Coneri}}, \bibinfo {author} {\bibfnamefont {S.}~\bibnamefont {Sanna}},
  \bibinfo {author} {\bibfnamefont {K.}~\bibnamefont {Zheng}}, \bibinfo
  {author} {\bibfnamefont {J.}~\bibnamefont {Lord}}, \ and\ \bibinfo {author}
  {\bibfnamefont {R.}~\bibnamefont {De~Renzi}},\ }\bibfield  {title} {\enquote
  {\bibinfo {title} {{Magnetic states of lightly hole-doped cuprates in the
  clean limit as seen via zero-field muon spin spectroscopy}},}\ }\href
  {\doibase 10.1103/PhysRevB.81.104507} {\bibfield  {journal} {\bibinfo
  {journal} {Phys. Rev. B}\ }\textbf {\bibinfo {volume} {81}},\ \bibinfo
  {pages} {104507} (\bibinfo {year} {2010})}\BibitemShut {NoStop}%
\bibitem [{\citenamefont {Blanco-Canosa}\ \emph {et~al.}(2014)\citenamefont
  {Blanco-Canosa}, \citenamefont {Frano}, \citenamefont {Schierle},
  \citenamefont {Porras}, \citenamefont {Loew}, \citenamefont {Minola},
  \citenamefont {Bluschke}, \citenamefont {Weschke}, \citenamefont {Keimer},\
  and\ \citenamefont {Le~Tacon}}]{Blanco-Canosa14_PRB}%
  \BibitemOpen
  \bibfield  {author} {\bibinfo {author} {\bibfnamefont {S.}~\bibnamefont
  {Blanco-Canosa}}, \bibinfo {author} {\bibfnamefont {A.}~\bibnamefont
  {Frano}}, \bibinfo {author} {\bibfnamefont {E.}~\bibnamefont {Schierle}},
  \bibinfo {author} {\bibfnamefont {J.}~\bibnamefont {Porras}}, \bibinfo
  {author} {\bibfnamefont {T.}~\bibnamefont {Loew}}, \bibinfo {author}
  {\bibfnamefont {M.}~\bibnamefont {Minola}}, \bibinfo {author} {\bibfnamefont
  {M.}~\bibnamefont {Bluschke}}, \bibinfo {author} {\bibfnamefont
  {E.}~\bibnamefont {Weschke}}, \bibinfo {author} {\bibfnamefont
  {B.}~\bibnamefont {Keimer}}, \ and\ \bibinfo {author} {\bibfnamefont
  {M.}~\bibnamefont {Le~Tacon}},\ }\bibfield  {title} {\enquote {\bibinfo
  {title} {{Resonant x-ray scattering study of charge-density wave correlations
  in ${\mathrm{YBa}}_{2}{\mathrm{Cu}}_{3}{\mathrm{O}}_{6+x}$}},}\ }\href
  {\doibase 10.1103/PhysRevB.90.054513} {\bibfield  {journal} {\bibinfo
  {journal} {Phys. Rev. B}\ }\textbf {\bibinfo {volume} {90}},\ \bibinfo
  {pages} {054513} (\bibinfo {year} {2014})}\BibitemShut {NoStop}%
\bibitem [{\citenamefont {Lei}\ \emph {et~al.}(1993)\citenamefont {Lei},
  \citenamefont {Sarrao}, \citenamefont {Visscher}, \citenamefont {Bell},
  \citenamefont {Thompson}, \citenamefont {Migliori}, \citenamefont {Welp},\
  and\ \citenamefont {Veal}}]{Lei93_PRB}%
  \BibitemOpen
  \bibfield  {author} {\bibinfo {author} {\bibfnamefont {M.}~\bibnamefont
  {Lei}}, \bibinfo {author} {\bibfnamefont {J.~L.}\ \bibnamefont {Sarrao}},
  \bibinfo {author} {\bibfnamefont {W.~M.}\ \bibnamefont {Visscher}}, \bibinfo
  {author} {\bibfnamefont {T.~M.}\ \bibnamefont {Bell}}, \bibinfo {author}
  {\bibfnamefont {J.~D.}\ \bibnamefont {Thompson}}, \bibinfo {author}
  {\bibfnamefont {A.}~\bibnamefont {Migliori}}, \bibinfo {author}
  {\bibfnamefont {U.~W.}\ \bibnamefont {Welp}}, \ and\ \bibinfo {author}
  {\bibfnamefont {B.~W.}\ \bibnamefont {Veal}},\ }\bibfield  {title} {\enquote
  {\bibinfo {title} {{Elastic constants of a monocrystal of superconducting
  ${\mathrm{YBa}}_{2}$${\mathrm{Cu}}_{3}$${\mathrm{O}}_{7\mathrm{\ensuremath{-}}\mathrm{\ensuremath{\delta}}}$}},}\
  }\href {\doibase 10.1103/PhysRevB.47.6154} {\bibfield  {journal} {\bibinfo
  {journal} {Phys. Rev. B}\ }\textbf {\bibinfo {volume} {47}},\ \bibinfo
  {pages} {6154--6156} (\bibinfo {year} {1993})}\BibitemShut {NoStop}%
\bibitem [{\citenamefont {Kim}\ \emph {et~al.}()\citenamefont {Kim},
  \citenamefont {Lefran{\c{c}}ois}, \citenamefont {Kummer}, \citenamefont
  {Fumagalli}, \citenamefont {Brookes}, \citenamefont {Betto}, \citenamefont
  {Nakata}, \citenamefont {Tortora}, \citenamefont {Porras}, \citenamefont
  {Loew}, \citenamefont {Barber}, \citenamefont {Braicovich}, \citenamefont
  {Ghiringhelli}, \citenamefont {Mackenzie}, \citenamefont {Hicks},
  \citenamefont {Keimer}, \citenamefont {Minola},\ and\ \citenamefont
  {Le~Tacon}}]{Kim_unpublished}%
  \BibitemOpen
  \bibfield  {author} {\bibinfo {author} {\bibfnamefont {H.-H.}\ \bibnamefont
  {Kim}}, \bibinfo {author} {\bibfnamefont {E.}~\bibnamefont
  {Lefran{\c{c}}ois}}, \bibinfo {author} {\bibfnamefont {K.}~\bibnamefont
  {Kummer}}, \bibinfo {author} {\bibfnamefont {R.}~\bibnamefont {Fumagalli}},
  \bibinfo {author} {\bibfnamefont {N.}~\bibnamefont {Brookes}}, \bibinfo
  {author} {\bibfnamefont {D.}~\bibnamefont {Betto}}, \bibinfo {author}
  {\bibfnamefont {S.}~\bibnamefont {Nakata}}, \bibinfo {author} {\bibfnamefont
  {M.}~\bibnamefont {Tortora}}, \bibinfo {author} {\bibfnamefont
  {J.}~\bibnamefont {Porras}}, \bibinfo {author} {\bibfnamefont
  {T.}~\bibnamefont {Loew}}, \bibinfo {author} {\bibfnamefont {M.~E.}\
  \bibnamefont {Barber}}, \bibinfo {author} {\bibfnamefont {L.}~\bibnamefont
  {Braicovich}}, \bibinfo {author} {\bibfnamefont {G.}~\bibnamefont
  {Ghiringhelli}}, \bibinfo {author} {\bibfnamefont {A.~P.}\ \bibnamefont
  {Mackenzie}}, \bibinfo {author} {\bibfnamefont {C.~W.}\ \bibnamefont
  {Hicks}}, \bibinfo {author} {\bibfnamefont {B.}~\bibnamefont {Keimer}},
  \bibinfo {author} {\bibfnamefont {M.}~\bibnamefont {Minola}}, \ and\ \bibinfo
  {author} {\bibfnamefont {M.}~\bibnamefont {Le~Tacon}},\ }\href@noop {}
  {\enquote {\bibinfo {title} {{Charge density waves in
  ${\mathrm{YBa}}_{2}{\mathrm{Cu}}_{3}{\mathrm{O}}_{6.67}$ probed by resonant
  x-ray scattering under uniaxial compression}},}\ }\bibinfo {note}
  {{unpublished}}\BibitemShut {NoStop}%
\bibitem [{\citenamefont {LeBoeuf}\ \emph {et~al.}(2013)\citenamefont
  {LeBoeuf}, \citenamefont {Kr{\"{a}}mer}, \citenamefont {Hardy}, \citenamefont
  {Liang}, \citenamefont {Bonn},\ and\ \citenamefont
  {Proust}}]{LeBoeuf13_NatPhys}%
  \BibitemOpen
  \bibfield  {author} {\bibinfo {author} {\bibfnamefont {D.}~\bibnamefont
  {LeBoeuf}}, \bibinfo {author} {\bibfnamefont {S.}~\bibnamefont
  {Kr{\"{a}}mer}}, \bibinfo {author} {\bibfnamefont {W.~N.}\ \bibnamefont
  {Hardy}}, \bibinfo {author} {\bibfnamefont {R.}~\bibnamefont {Liang}},
  \bibinfo {author} {\bibfnamefont {D.~A.}\ \bibnamefont {Bonn}}, \ and\
  \bibinfo {author} {\bibfnamefont {C.}~\bibnamefont {Proust}},\ }\bibfield
  {title} {\enquote {\bibinfo {title} {{Thermodynamic phase diagram of static
  charge order in underdoped
  ${\mathrm{YBa}}_{2}$${\mathrm{Cu}}_{3}$${\mathrm{O}}_{y}$}},}\ }\href
  {\doibase 10.1038/nphys2502} {\bibfield  {journal} {\bibinfo  {journal} {Nat.
  Physics}\ }\textbf {\bibinfo {volume} {9}},\ \bibinfo {pages} {79--83}
  (\bibinfo {year} {2013})}\BibitemShut {NoStop}%
\bibitem [{\citenamefont {Jang}\ \emph {et~al.}(2016)\citenamefont {Jang},
  \citenamefont {Lee}, \citenamefont {Nojiri}, \citenamefont {Matsuzawa},
  \citenamefont {Yasumura}, \citenamefont {Nie}, \citenamefont {Maharaj},
  \citenamefont {Gerber}, \citenamefont {Liu}, \citenamefont {Mehta},
  \citenamefont {Bonn}, \citenamefont {Liang}, \citenamefont {Hardy},
  \citenamefont {Burns}, \citenamefont {Islam}, \citenamefont {Song},
  \citenamefont {Hastings}, \citenamefont {Devereaux}, \citenamefont {Shen},
  \citenamefont {Kivelson}, \citenamefont {Kao}, \citenamefont {Zhu},\ and\
  \citenamefont {Lee}}]{Jang16_PNAS}%
  \BibitemOpen
  \bibfield  {author} {\bibinfo {author} {\bibfnamefont {H.}~\bibnamefont
  {Jang}}, \bibinfo {author} {\bibfnamefont {W.-S.}\ \bibnamefont {Lee}},
  \bibinfo {author} {\bibfnamefont {H.}~\bibnamefont {Nojiri}}, \bibinfo
  {author} {\bibfnamefont {S.}~\bibnamefont {Matsuzawa}}, \bibinfo {author}
  {\bibfnamefont {H.}~\bibnamefont {Yasumura}}, \bibinfo {author}
  {\bibfnamefont {L.}~\bibnamefont {Nie}}, \bibinfo {author} {\bibfnamefont
  {A.~V.}\ \bibnamefont {Maharaj}}, \bibinfo {author} {\bibfnamefont
  {S.}~\bibnamefont {Gerber}}, \bibinfo {author} {\bibfnamefont {Y.-J.}\
  \bibnamefont {Liu}}, \bibinfo {author} {\bibfnamefont {A.}~\bibnamefont
  {Mehta}}, \bibinfo {author} {\bibfnamefont {D.~A.}\ \bibnamefont {Bonn}},
  \bibinfo {author} {\bibfnamefont {R.}~\bibnamefont {Liang}}, \bibinfo
  {author} {\bibfnamefont {W.~N.}\ \bibnamefont {Hardy}}, \bibinfo {author}
  {\bibfnamefont {C.~A.}\ \bibnamefont {Burns}}, \bibinfo {author}
  {\bibfnamefont {Z.}~\bibnamefont {Islam}}, \bibinfo {author} {\bibfnamefont
  {S.}~\bibnamefont {Song}}, \bibinfo {author} {\bibfnamefont {J.}~\bibnamefont
  {Hastings}}, \bibinfo {author} {\bibfnamefont {T.~P.}\ \bibnamefont
  {Devereaux}}, \bibinfo {author} {\bibfnamefont {Z.-X.}\ \bibnamefont {Shen}},
  \bibinfo {author} {\bibfnamefont {S.~A.}\ \bibnamefont {Kivelson}}, \bibinfo
  {author} {\bibfnamefont {C.-C.}\ \bibnamefont {Kao}}, \bibinfo {author}
  {\bibfnamefont {D.}~\bibnamefont {Zhu}}, \ and\ \bibinfo {author}
  {\bibfnamefont {J.-S.}\ \bibnamefont {Lee}},\ }\bibfield  {title} {\enquote
  {\bibinfo {title} {{Ideal charge-density-wave order in the high-field state
  of superconducting YBCO}},}\ }\href {\doibase 10.1073/pnas.1612849113}
  {\bibfield  {journal} {\bibinfo  {journal} {Proc. Natl. Acad. Sci. U.S.A.}\
  }\textbf {\bibinfo {volume} {113}},\ \bibinfo {pages} {14645--14650}
  (\bibinfo {year} {2016})}\BibitemShut {NoStop}%
\bibitem [{\citenamefont {Corboz}\ \emph {et~al.}(2014)\citenamefont {Corboz},
  \citenamefont {Rice},\ and\ \citenamefont {Troyer}}]{Corboz14_PRL}%
  \BibitemOpen
  \bibfield  {author} {\bibinfo {author} {\bibfnamefont {P.}~\bibnamefont
  {Corboz}}, \bibinfo {author} {\bibfnamefont {T.~M.}\ \bibnamefont {Rice}}, \
  and\ \bibinfo {author} {\bibfnamefont {M.}~\bibnamefont {Troyer}},\
  }\bibfield  {title} {\enquote {\bibinfo {title} {{Competing States in the
  $t$-$J$ Model: Uniform $d$-Wave State versus Stripe State}},}\ }\href
  {\doibase 10.1103/PhysRevLett.113.046402} {\bibfield  {journal} {\bibinfo
  {journal} {Phys. Rev. Lett.}\ }\textbf {\bibinfo {volume} {113}},\ \bibinfo
  {pages} {046402} (\bibinfo {year} {2014})}\BibitemShut {NoStop}%
\bibitem [{\citenamefont {H{\"u}cker}\ \emph {et~al.}(2011)\citenamefont
  {H{\"u}cker}, \citenamefont {v.~Zimmermann}, \citenamefont {Gu},
  \citenamefont {Xu}, \citenamefont {Wen}, \citenamefont {Xu}, \citenamefont
  {Kang}, \citenamefont {Zheludev},\ and\ \citenamefont
  {Tranquada}}]{Huecker11_PRB}%
  \BibitemOpen
  \bibfield  {author} {\bibinfo {author} {\bibfnamefont {M.}~\bibnamefont
  {H{\"u}cker}}, \bibinfo {author} {\bibfnamefont {M.}~\bibnamefont
  {v.~Zimmermann}}, \bibinfo {author} {\bibfnamefont {G.~D.}\ \bibnamefont
  {Gu}}, \bibinfo {author} {\bibfnamefont {Z.~J.}\ \bibnamefont {Xu}}, \bibinfo
  {author} {\bibfnamefont {J.~S.}\ \bibnamefont {Wen}}, \bibinfo {author}
  {\bibfnamefont {G.}~\bibnamefont {Xu}}, \bibinfo {author} {\bibfnamefont
  {H.~J.}\ \bibnamefont {Kang}}, \bibinfo {author} {\bibfnamefont
  {A.}~\bibnamefont {Zheludev}}, \ and\ \bibinfo {author} {\bibfnamefont
  {J.~M.}\ \bibnamefont {Tranquada}},\ }\bibfield  {title} {\enquote {\bibinfo
  {title} {{Stripe order in superconducting
  La${}_{2\ensuremath{-}x}$Ba${}_{x}$CuO${}_{4}$
  ($0.095\ensuremath{\leqslant}x\ensuremath{\leqslant}0.155$)}},}\ }\href
  {\doibase 10.1103/PhysRevB.83.104506} {\bibfield  {journal} {\bibinfo
  {journal} {Phys. Rev. B}\ }\textbf {\bibinfo {volume} {83}},\ \bibinfo
  {pages} {104506} (\bibinfo {year} {2011})}\BibitemShut {NoStop}%
\bibitem [{\citenamefont {Guguchia}\ \emph {et~al.}(2020)\citenamefont
  {Guguchia}, \citenamefont {Das}, \citenamefont {Wang}, \citenamefont
  {Adachi}, \citenamefont {Kitajima}, \citenamefont {Elender}, \citenamefont
  {Br{\"{u}}ckner}, \citenamefont {Ghosh}, \citenamefont {Grinenko},
  \citenamefont {Shiroka}, \citenamefont {M{\"{u}}ller}, \citenamefont {Mudry},
  \citenamefont {Baines}, \citenamefont {Bartkowiak}, \citenamefont {Koike},
  \citenamefont {Amato}, \citenamefont {Tranquada}, \citenamefont {Klauss},
  \citenamefont {Hicks},\ and\ \citenamefont {Luetkens}}]{Guguchia20_PRL}%
  \BibitemOpen
  \bibfield  {author} {\bibinfo {author} {\bibfnamefont {Z.}~\bibnamefont
  {Guguchia}}, \bibinfo {author} {\bibfnamefont {D.}~\bibnamefont {Das}},
  \bibinfo {author} {\bibfnamefont {C.~N.}\ \bibnamefont {Wang}}, \bibinfo
  {author} {\bibfnamefont {T.}~\bibnamefont {Adachi}}, \bibinfo {author}
  {\bibfnamefont {N.}~\bibnamefont {Kitajima}}, \bibinfo {author}
  {\bibfnamefont {M.}~\bibnamefont {Elender}}, \bibinfo {author} {\bibfnamefont
  {F.}~\bibnamefont {Br{\"{u}}ckner}}, \bibinfo {author} {\bibfnamefont
  {S.}~\bibnamefont {Ghosh}}, \bibinfo {author} {\bibfnamefont
  {V.}~\bibnamefont {Grinenko}}, \bibinfo {author} {\bibfnamefont
  {T.}~\bibnamefont {Shiroka}}, \bibinfo {author} {\bibfnamefont
  {M.}~\bibnamefont {M{\"{u}}ller}}, \bibinfo {author} {\bibfnamefont
  {C.}~\bibnamefont {Mudry}}, \bibinfo {author} {\bibfnamefont
  {C.}~\bibnamefont {Baines}}, \bibinfo {author} {\bibfnamefont
  {M.}~\bibnamefont {Bartkowiak}}, \bibinfo {author} {\bibfnamefont
  {Y.}~\bibnamefont {Koike}}, \bibinfo {author} {\bibfnamefont
  {A.}~\bibnamefont {Amato}}, \bibinfo {author} {\bibfnamefont {J.~M.}\
  \bibnamefont {Tranquada}}, \bibinfo {author} {\bibfnamefont {H.-H.}\
  \bibnamefont {Klauss}}, \bibinfo {author} {\bibfnamefont {C.~W.}\
  \bibnamefont {Hicks}}, \ and\ \bibinfo {author} {\bibfnamefont
  {H.}~\bibnamefont {Luetkens}},\ }\bibfield  {title} {\enquote {\bibinfo
  {title} {{Using Uniaxial Stress to Probe the Relationship between Competing
  Superconducting States in a Cuprate with Spin-stripe Order}},}\ }\href
  {\doibase 10.1103/PhysRevLett.125.097005} {\bibfield  {journal} {\bibinfo
  {journal} {Phys. Rev. Lett.}\ }\textbf {\bibinfo {volume} {125}},\ \bibinfo
  {pages} {097005} (\bibinfo {year} {2020})}\BibitemShut {NoStop}%
\bibitem [{\citenamefont {Ikeda}\ \emph {et~al.}(2019)\citenamefont {Ikeda},
  \citenamefont {Straquadine}, \citenamefont {Hristov}, \citenamefont
  {Worasaran}, \citenamefont {Palmstrom}, \citenamefont {Sorensen},
  \citenamefont {Walmsley},\ and\ \citenamefont {Fisher}}]{Ikeda19_RSI}%
  \BibitemOpen
  \bibfield  {author} {\bibinfo {author} {\bibfnamefont {M.~S.}\ \bibnamefont
  {Ikeda}}, \bibinfo {author} {\bibfnamefont {J.~A.~W.}\ \bibnamefont
  {Straquadine}}, \bibinfo {author} {\bibfnamefont {A.~T.}\ \bibnamefont
  {Hristov}}, \bibinfo {author} {\bibfnamefont {T.}~\bibnamefont {Worasaran}},
  \bibinfo {author} {\bibfnamefont {J.~C.}\ \bibnamefont {Palmstrom}}, \bibinfo
  {author} {\bibfnamefont {M.}~\bibnamefont {Sorensen}}, \bibinfo {author}
  {\bibfnamefont {P.}~\bibnamefont {Walmsley}}, \ and\ \bibinfo {author}
  {\bibfnamefont {I.~R.}\ \bibnamefont {Fisher}},\ }\bibfield  {title}
  {\enquote {\bibinfo {title} {{AC elastocaloric effect as a probe for
  thermodynamic signatures of continuous phase transitions}},}\ }\href
  {\doibase 10.1063/1.5099924} {\bibfield  {journal} {\bibinfo  {journal} {Rev.
  Sci. Instrum.}\ }\textbf {\bibinfo {volume} {90}},\ \bibinfo {pages} {083902}
  (\bibinfo {year} {2019})}\BibitemShut {NoStop}%
\bibitem [{\citenamefont {Zhang}\ \emph {et~al.}(2017)\citenamefont {Zhang},
  \citenamefont {Levenson-Falk}, \citenamefont {Ramshaw}, \citenamefont {Bonn},
  \citenamefont {Liang}, \citenamefont {Hardy}, \citenamefont {Hartnoll},\ and\
  \citenamefont {Kapitulnik}}]{Zhang17_PNAS}%
  \BibitemOpen
  \bibfield  {author} {\bibinfo {author} {\bibfnamefont {J.}~\bibnamefont
  {Zhang}}, \bibinfo {author} {\bibfnamefont {E.~M.}\ \bibnamefont
  {Levenson-Falk}}, \bibinfo {author} {\bibfnamefont {B.~J.}\ \bibnamefont
  {Ramshaw}}, \bibinfo {author} {\bibfnamefont {D.~A.}\ \bibnamefont {Bonn}},
  \bibinfo {author} {\bibfnamefont {R.}~\bibnamefont {Liang}}, \bibinfo
  {author} {\bibfnamefont {W.~N.}\ \bibnamefont {Hardy}}, \bibinfo {author}
  {\bibfnamefont {S.~A.}\ \bibnamefont {Hartnoll}}, \ and\ \bibinfo {author}
  {\bibfnamefont {A.}~\bibnamefont {Kapitulnik}},\ }\bibfield  {title}
  {\enquote {\bibinfo {title} {{Anomalous thermal diffusivity in underdoped
  YBa$_2$Cu$_3$O$_{6+x}$}},}\ }\href {\doibase 10.1073/pnas.1703416114}
  {\bibfield  {journal} {\bibinfo  {journal} {Proc. Natl. Acad. Sci. U.S.A.}\
  }\textbf {\bibinfo {volume} {114}},\ \bibinfo {pages} {5378--5383} (\bibinfo
  {year} {2017})}\BibitemShut {NoStop}%
\bibitem [{\citenamefont {Loram}\ \emph {et~al.}(1993)\citenamefont {Loram},
  \citenamefont {Mirza}, \citenamefont {Cooper},\ and\ \citenamefont
  {Liang}}]{Loram93_PRL}%
  \BibitemOpen
  \bibfield  {author} {\bibinfo {author} {\bibfnamefont {J.~W.}\ \bibnamefont
  {Loram}}, \bibinfo {author} {\bibfnamefont {K.~A.}\ \bibnamefont {Mirza}},
  \bibinfo {author} {\bibfnamefont {J.~R.}\ \bibnamefont {Cooper}}, \ and\
  \bibinfo {author} {\bibfnamefont {W.~Y.}\ \bibnamefont {Liang}},\ }\bibfield
  {title} {\enquote {\bibinfo {title} {{Electronic specific heat of
  ${\mathrm{YBa}}_{2}$${\mathrm{Cu}}_{3}$${\mathrm{O}}_{6+\mathit{x}}$ from 1.8
  to 300 K}},}\ }\href {\doibase 10.1103/PhysRevLett.71.1740} {\bibfield
  {journal} {\bibinfo  {journal} {Phys. Rev. Lett.}\ }\textbf {\bibinfo
  {volume} {71}},\ \bibinfo {pages} {1740--1743} (\bibinfo {year}
  {1993})}\BibitemShut {NoStop}%
\end{thebibliography}
\end{document}

\title{Supplemental Materials for:\texorpdfstring{\\}{}Suppression of superconductivity by charge density wave order in \texorpdfstring{YBa\textsubscript{2}Cu\textsubscript{3}O\textsubscript{6.67}}{YBa2Cu3O6.67}} 

\author{Mark E. Barber}
\altaffiliation[Present address: ]{Department of Applied Physics and Geballe Laboratory for Advanced Materials, Stanford University, Stanford, CA 94305} 
\email{mebarber@stanford.edu}
\affiliation{Max Planck Institute for Chemical Physics of Solids, N{\"o}thnitzer Stra{\ss}e 40, 01187 Dresden, Germany}
\author{Hun-ho Kim}
\author{Toshinao Loew}
\affiliation{Max Planck Institute for Solid State Research, Heisenbergstra{\ss}e 1, 70569 Stuttgart, Germany}
\author{Matthieu Le Tacon}
\affiliation{Max Planck Institute for Solid State Research, Heisenbergstra{\ss}e 1, 70569 Stuttgart, Germany}
\affiliation{Karlsruhe Institute of Technology, Institute for Quantum Materials and Technologies, Hermann-von-Helmholtz-Platz 1, 76344 Eggenstein-Leopoldshafen, Germany}
\author{Matteo Minola}
\affiliation{Max Planck Institute for Solid State Research, Heisenbergstra{\ss}e 1, 70569 Stuttgart, Germany}
\author{Marcin Konczykowski}
\affiliation{Laboratoire des Solides Irradi{\'e}s, CEA/DRF/lRAMIS, Ecole Polytechnique, CNRS, Institut Polytechnique de Paris, F-91128 Palaiseau, France}
\author{Bernhard Keimer}
\affiliation{Max Planck Institute for Solid State Research, Heisenbergstra{\ss}e 1, 70569 Stuttgart, Germany}
\author{Andrew P. Mackenzie}
\affiliation{Max Planck Institute for Chemical Physics of Solids, N{\"o}thnitzer Stra{\ss}e 40, 01187 Dresden, Germany}
\affiliation{Scottish Universities Physics Alliance, School of Physics and Astronomy, University of St.\ Andrews, St.\ Andrews KY16 9SS, U.K.}
\author{Clifford W. Hicks} 
\email{hicks@cpfs.mpg.de}
\affiliation{Max Planck Institute for Chemical Physics of Solids, N{\"o}thnitzer Stra{\ss}e 40, 01187 Dresden, Germany}
\affiliation{School of Physics and Astronomy, University of Birmingham, Birmingham B15 2TT, U.K.}

\date{}

\maketitle

\section{I.\texorpdfstring{\quad}{} More information on the experiment setup}
Samples were grown using a flux method~\cite{Lin92_PhysicaC} producing large single crystals whose oxygen content was later adjusted to 6.67 ($T_{\textrm{c}} \approx 65$~K) by annealing under well-defined oxygen partial pressure.
The large samples were mechanically detwinned by heating under slight uniaxial stress, 50--60~MPa, to 400\,$^{\circ}$C, before being cut into smaller pieces and mechanically polished to the required dimensions for the pressure cell.
Samples were mounted into the uniaxial stress cell with Stycast 2850FT epoxy, cured for 4 hours at 65\,$^{\circ}$C.
The epoxy layers on both the top and bottom faces of the sample ends were $\approx$30~$\mu$m thick.

The uniaxial pressure cell uses three piezoelectric actuators, arranged to cancel their own differential thermal expansion, to apply pressure to the sample, which is shaped into a long, thin bar.
Making reference to Fig.~1(a), one end of the sample is attached to moving block A and the other to moving block B.
These blocks are joined with flexures to the outer frame of the cell.
The actuators drive motion of moving block A, which is constrained to move along a straight line by four flexures.
Block B is held by four thicker flexures, and moves slightly in response to the force transmitted through the sample.
A parallel-plate capacitive sensor measures the displacement of block B, and combining with the known spring constant of the thick flexures the force on the sample is determined.
A complete description of the apparatus and its calibration is given in Ref.~\cite{Barber19_RSI}.

\begin{figure}[htp]
 \centering
 \includegraphics{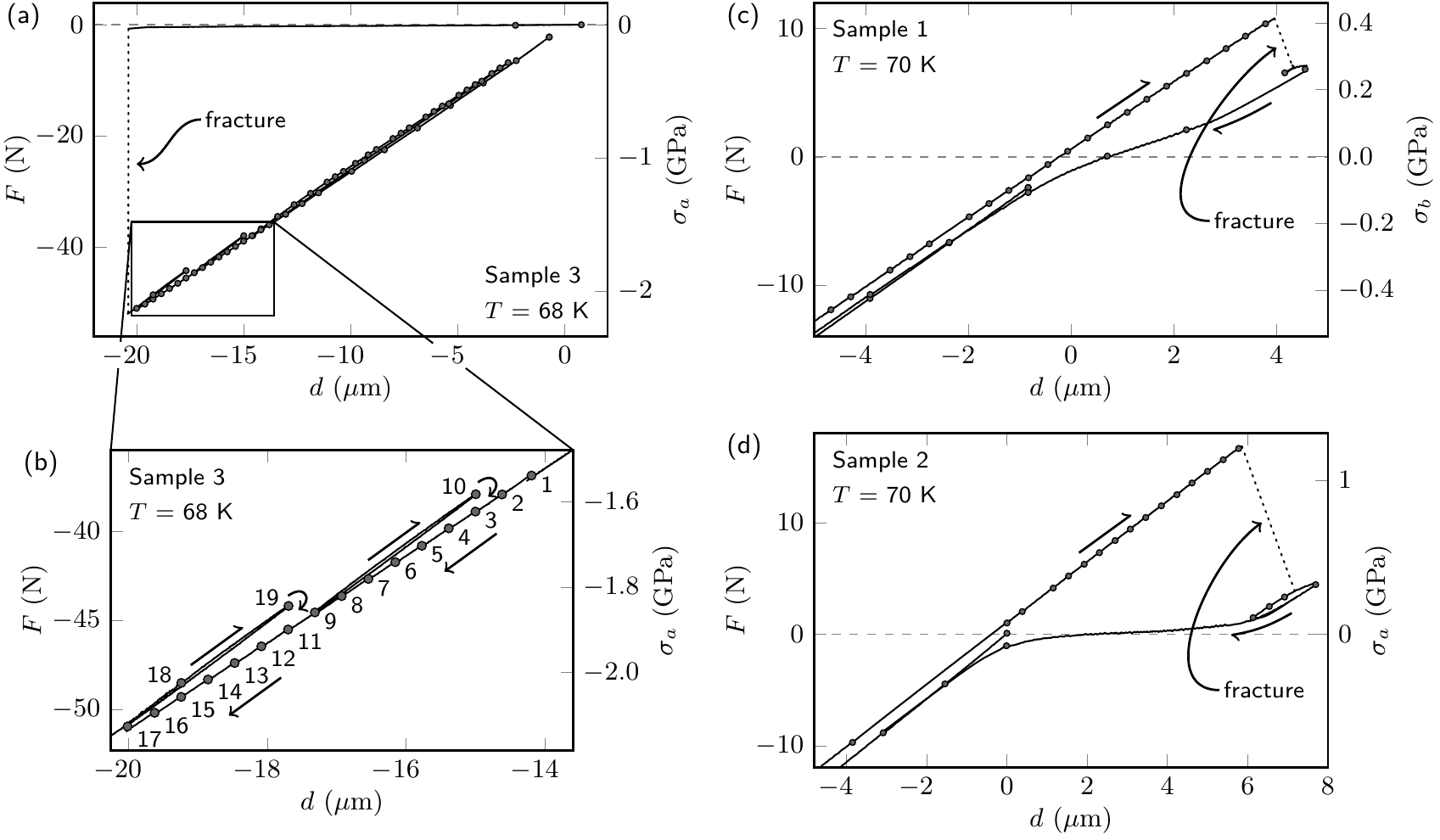}
 \caption{\label{figS1}(a)~Force-displacement data for a series of $T_{\mathrm{c}}$ measurements for Sample 3.
 The markers indicate points where the temperature was ramped from 68~K to below $T_{\mathrm{c}}$ and then back to 68~K at fixed strain.
 Between these points the displacement $d$ was ramped at 1.5~nm/s.
(b)~Close-up of the data in panel (a), showing the probable plastic relaxation of the epoxy holding the sample as $|\sigma_a|$ became large.
 The numbers on each marker indicate the order in which temperature ramps were performed.
(c)~Force-displacement data for Sample 1, before and after it was fractured under tension.
 As in panel (a), the markers indicate strains where temperature ramps were performed to measure $T_{\mathrm{c}}$, and the lines are the strain ramps between these points.
 The displacement was ramped at 1.5~nm/s.
(d)~Same as (c), for Sample 2.
}
\end{figure}

\begin{figure}[htp]
 \centering
 \includegraphics{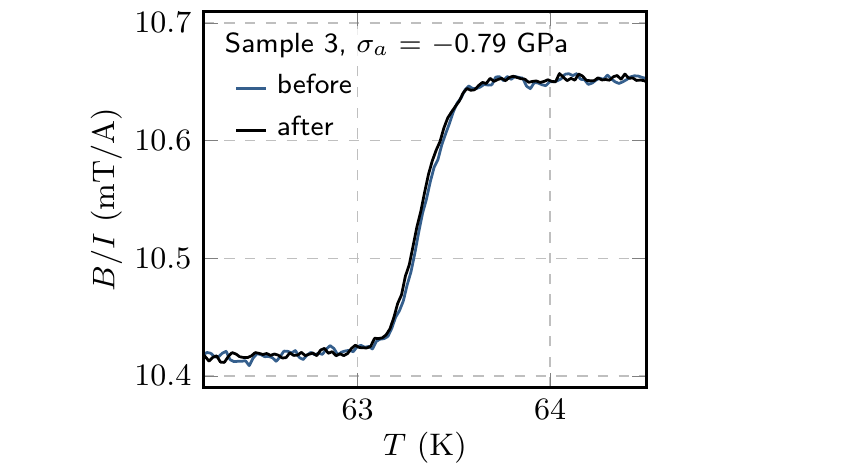}
 \caption{\label{figS2}Two measurements of Sample 3 at $\sigma_a = -$0.79~GPa, before and after $\sigma_{a}$ was ramped to $-$1.6~GPa.
}
\end{figure}

There is in addition a capacitive displacement sensor between moving blocks A and B, which measures the displacement applied to the sample and the epoxy that holds it.
Data from this sensor provides information on the mechanical state of the sample and epoxy.
Force-displacement data from Sample 3, as the applied stress approached $-2$~GPa at $T = 68$~K, are shown in Fig.~\ref{figS1}(a--b).
Panel (a) shows $F(d)$, where $F$ is force and $d$ displacement, over the entire course of measurements, including the point where the sample fractured under compression and $F$ dropped abruptly.
Panel (b) shows a close-up of $F(d)$ at high stresses, approaching the point of fracture.
Here, it is seen that $F(d)$ followed a gentler slope as $|\sigma_a|$ was increased than on the return strokes.
This difference is visible in, for example, the gentler slope between the points numbered 1 and 9 in the figure, where $|\sigma_{a}|$ was increased monotonically, and the steeper slope between points 9 and 10, where $|\sigma_{a}|$ was decreased.
This behaviour most likely shows that the epoxy relaxes plastically as high stresses are approached.
That this possible plastic flow is in the epoxy, not the sample, is shown by the data in Fig.~\ref{figS2}, where the transition observed in Sample 3 at $-0.79$~GPa is the same before and after ramping to $-1.6$~GPa.

We note that this smooth plastic relaxation of the epoxy is in contrast to the sharp fractures observed at 5~K in Ref.~\cite{Barber19_RSI}.
The sample mounting procedure was the same, so we hypothesise that the different behaviour is due to the temperature difference.

The force sensor reading varied by $\sim$2~N from cool-down to cool-down, and so its zero-force reading was determined in situ by deliberately fracturing each sample after measurement, on the theory that with the sample broken, the applied force would necessarily be zero.
For Sample 3, the fracture was complete --- a portion of the sample disappeared and the link between the two ends was completely broken --- and so the $F=0$ reading could be determined with high precision.
Samples 1 and 2, on the other hand, were fractured in tension, and the fractures occurred inside the epoxy, such that there remained a frictional connection between the two sample ends.
$F(d)$ traces leading up to and following the fractures for Samples 1 and 2 are shown in Figs.~\ref{figS1}(c) and (d), respectively.
For both samples, after fracture there is a portion of the force-displacement curve where the slope is strongly reduced, which we interpret as the crossover from tensile to compressive stress in the sample, broadened by frictional and plastic effects in the epoxy.
Because there is no single sharp feature that can be identified as zero stress, we estimate an error of $\pm 0.05$~GPa on the zero-stress determination for these two samples. 

The burn-through field for all measurements with the Hall cross susceptometer shown in this paper, apart from Fig.~2(a), was $\sim$200~$\mu$T at 20~kHz.
The probing field from the susceptometer was $\sim$25~$\mu$T, applied at 211~Hz, for all measurements.\\

\section{II.\texorpdfstring{\quad}{} Additional data}
\begin{figure}[hb]
 \centering
 \includegraphics{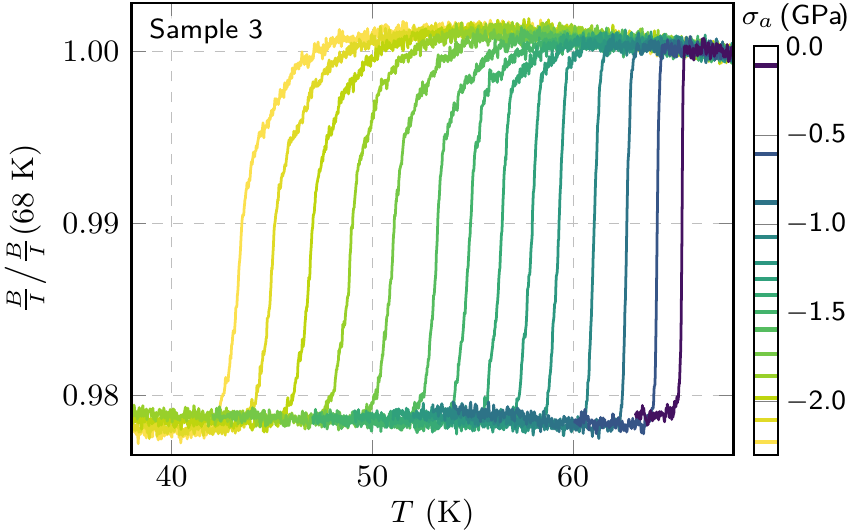}
 \caption{\label{figS3}Real part of the susceptometer response $B/I$ for Sample 3 at various $\sigma_{a}$.
 Due to drift, data are scaled by the readings at 68~K.
}
\end{figure}

Susceptibility data across $T_{\mathrm{c}}$ for Sample 3 are shown in Fig.~\ref{figS3}.
(Susceptibility data from Samples 1 and 2 are shown in Fig.~2.)
For this set of measurements, the signal from the Hall cross drifted over time, so data are scaled by dividing each sweep by the starting value at 68~K.\\

\section{III.\texorpdfstring{\quad}{} Further discussion of \texorpdfstring{{\boldmath $\sigma_{\mathrm{CDW}}$}}{sigma\_CDW}}
As noted in the main text, in Ref.~[6] the strain at which 3D CDW order onsets in \YBCO{} was reported to be between $-0.8$ and $-1.0 \cdot 10^{-2}$, corresponding to $\sigma_{\mathrm{CDW}}$ between $-1.3$ and $-1.6$~GPa.
This is somewhat larger than the $\sigma_{\mathrm{CDW}}$ found here for Samples 2 and 3, respectively $-0.97$ and $-1.2$~GPa.
However, the stress apparatus used in Ref.~[6] incorporated a sensor only of the applied displacement, not the applied force.
A conversion factor between displacement and strain was obtained using Bragg reflection data at low applied displacements.
At large displacements, the Bragg peak location began to saturate, and the origin of this behaviour was not understood; bending of the sample at large strains is one possibility.
Interpreting the large-strain Bragg data literally indicates $\sigma_{\mathrm{CDW}} = -1.0$~GPa, in good agreement with the data here.
Overall, within experimental accuracy so far and possible sample-to-sample variability, Ref.~[6] and the data here are in good agreement on $\sigma_{\mathrm{CDW}}$.\\

\section{IV.\texorpdfstring{\quad}{} Further information on the elastocaloric effect measurements}
Elastocaloric effect (ECE) measurements are ideally performed in the adiabatic limit, meaning that the measurement frequency should be high enough that temperature oscillations do not dissipate into the bath through the ends of the sample, yet low enough that the thermocouple thermalises to the sample.
Here, this frequency window was not broad.
In analysis below, we show that at our measurement frequency of 23.11~Hz the observed signal was about 80\% of its adiabatic limit.
This factor is included in the normalised ECE data of Fig.~5, and the good agreement with the heat capacity jump at $T_{\mathrm{c}}$, $\Delta C$, measured directly in Ref.~[36] shows that it is at least approximately correct.
The narrowness of the frequency window prevents us from commenting on small apparent changes in $\Delta C$, but our data are sufficient to conclude that $\Delta C$ does not change drastically across $\sigma_{\mathrm{CDW}}$.

\begin{figure}[tp]
 \centering
 \includegraphics{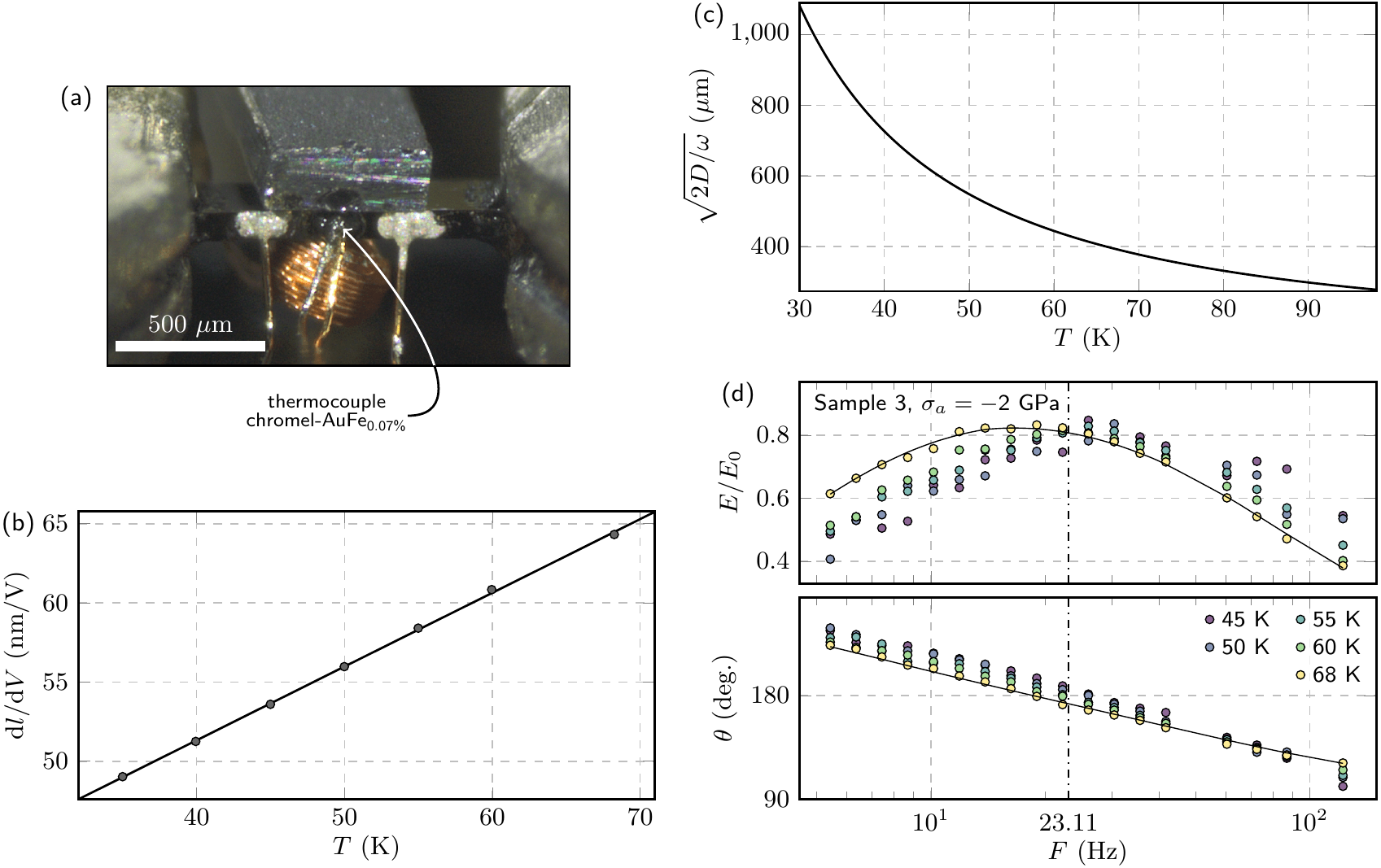}
 \caption{\label{figS4}(a)~Micrograph of Sample 3 showing the chromel-AuFe\textsubscript{0.07\%} thermocouple.
 (b)~Temperature dependence of the piezoelectric actuator response to applied voltage and a linear fit to the data.
 (c)~Temperature dependence of the characteristic thermal length of \YBCO{} at 23.1~Hz~[35].
 (d)~Frequency dependence of the elastocaloric effect signal, amplitude and phase, at various temperatures at $\sigma_a = -$2~GPa.
 The solid line is the fitted thermal transfer function at 68~K.
}
\end{figure}

We now provide more details on the measurement.
A photograph of Sample 3, including the attached thermocouple, is shown in Fig.~\ref{figS4}(a).
The thermocouple is chromel-AuFe\textsubscript{0.07\%}, and was secured to the sample with Stycast 2850FT.
It was made from wire from the same spools as in Ref.~\cite{Stockert11_Cryogenics}, and we therefore apply the calibration reported there.
The stress apparatus contains two sets of actuators, labelled compression and tension actuators, whose actions on the sample have opposite sign.
For the ECE measurements, the compression actuators were used to apply the dc stress, and the tension actuators the ac stress, with a temperature-independent applied voltage of 1.4~V\textsubscript{rms}.
The capacitance bridge monitoring the force sensor did not have enough bandwidth for measurements at 23~Hz, and we therefore obtain a dc calibration of the tension actuator response, and, following Refs.~[34] and \cite{PI}, make the approximation that this calibration applies at 23~Hz.
For Sample~3, at $\sigma_{a} \sim 0$ and $T = 68$~K the stress-voltage response was 4.9~MPa/V.
After the sample was fractured, we then tested the temperature response of the tension actuator by ramping the applied voltage between $-10$ and $+10$~V, yielding the data plotted in Fig.~\ref{figS4}(b).
Between 69 and 45~K, the actuator response decreases linearly with temperature by about 16\%.
We therefore take our applied ac stress to be 4.9~MPa/V $\times$ 1.4~V\textsubscript{rms} = 6.9~MPa\textsubscript{rms} at 68~K, and for lower temperatures scale this by the response curve shown in Fig.~\ref{figS4}(b).  

ECE measurements yield an elastocaloric coefficient $E \equiv \mathrm{d}T/\mathrm{d}\sigma$.
We now discuss analysis to determine $E/E_0$, where $E_0$ is the elastocaloric coefficient in the adiabatic limit.
Ikeda \textit{et al}.~[34] showed that the frequency dependence of the elastocaloric response can be well captured by a discretized thermal model.
In this model,
\begin{eqnarray*}
\frac{E}{E_0} & = & \left(a^2 + b^2\right)^{-1/2} \, , \\
\theta & = & \arctan(\frac{a}{b})\, , 
\end{eqnarray*}
where $\theta$ is the phase of the signal relative to the stress oscillation, and
\begin{eqnarray*}
a & = & \frac{1}{\omega \tau_i} - \omega \tau_\theta \, , \\
b & = & 1 + \frac{C_\theta}{C_\mathrm{s}} + \frac{\tau_\theta}{\tau_i} \, .
\end{eqnarray*}
$\tau_i$ is the time constant for thermalisation of the sample to the bath (approximated as being frequency-independent), $\tau_\theta$ is that for thermalisation of the temperature sensor to the sample, $C_\theta$ is the heat capacity of the sensor, and $C_\mathrm{s}$ is the heat capacity of the sample.
We estimate $C_\theta / C_{\mathrm{s}}$, and leave $\tau_i$ and $\tau_\theta$ as fitting parameters.
We estimate $C_\theta$ as the heat capacity of 700~$\mu$m of 25~$\mu$m-diameter chromel wire and 1~mm of 25~$\mu$m-diameter AuFe\textsubscript{0.07\%} wire, where the lengths are determined from the thermal diffusivities of these materials at 23~Hz and 65~K, in addition to a volume $150 \times 100 \times 90$~$\mu$m$^3$ of Stycast 2850FT.
At $T = (45, 50, 55, 60, 68)$~K, we find $C_\theta \sim (0.99, 1.1, 1.3, 1.4, 1.6)$~$\mu$J/K, and employ these values for the fitting.
$C_{\mathrm{s}}$ is much larger than $C_\theta$, and we therefore take its frequency dependence into account.
Following Ref.~[34], we take an approximate model in which at low frequencies $C_{\mathrm{s}}$ is the heat capacity of the entire exposed portion of the sample, and at high frequencies it is the heat capacity of a portion of the sample of length twice the thermal diffusion length $\xi$:
\begin{equation*}
C_{\mathrm{s}}  = \left\{ \begin{aligned} C A_{\mathrm{s}} l_{\mathrm{exp}} \, , & \hspace{1cm} & \omega < \omega_\text{1D} \, , \\ 2 C A_{\mathrm{s}} \xi \, , & \hspace{1cm} & \omega > \omega_\text{1D} \, ,
\end{aligned} \right.
\end{equation*}
where $C$ is the specific heat capacity of \YBCO{}, which we take from Ref.~[36], $A_{\mathrm{s}}$ is the cross-sectional area of the sample, and $l_{\mathrm{exp}} = 1$~mm is the exposed length of the sample.
$\xi = (2D/\omega)^{1/2}$, where $D$ is the thermal diffusivity, which we take from Ref.~[35].
$\omega_\text{1D}$ is the angular frequency at which $2\xi$ crosses $l_{\mathrm{exp}}$.
In Fig.~\ref{figS4}(c) we show the temperature dependence of $\xi$.
At our measurement frequency of 23.11~Hz, we find $C_\theta / C_{\mathrm{s}} \sim 0.1$ at each temperature studied here.

$\tau_i$ and $\tau_\theta$ are obtained from simultaneous fitting of the observed frequency dependence of $E$ and $\theta$ at $\sigma_a = -2.0$~GPa, where the elastocaloric response was measured against frequency.
Data are shown in Fig.~\ref{figS4}(d), along with the fitted thermal transfer function at 68~K.
In the graph, $E$ is scaled by $E_0$, which is obtained from independent fitting at each temperature.
The measurement frequency, 23.11~Hz, was selected as the frequency where $E/E_0$ is close to its maximum over the entire temperature range studied here.
Based on a linear fit of $E/E_0$ with temperature, we find $E/E_0$ to vary between 0.796 at 45~K and 0.810 at 68~K.
We apply this $T$-dependent correction to the data in Fig.~5 in the main text.
Because $C_\theta$ is small, a large change in our estimate for $C_\theta$ has minimal effect.
For example, doubling $C_\theta$ would reduce the maximum $E/E_0$ in the fitting to $\sim$0.7.\\